\documentclass[reprint, superscriptaddress,amsmath,amssymb,aps, prb]{revtex4-2}
\usepackage[dvipsnames]{xcolor}
\usepackage{graphicx}% Include figure files
\usepackage{dcolumn}% Align table columns on decimal point
\usepackage{bm}% bold math
\usepackage[colorlinks, breaklinks, linkcolor=green, %OrangeRed,
citecolor=RoyalBlue, urlcolor=NavyBlue]{hyperref}
\usepackage{siunitx}
\usepackage[normalem]{ulem}
\usepackage{accents}

\usepackage{nicefrac}
\usepackage{xspace}
\usepackage{comment}
\usepackage{layouts}
\usepackage{float}
\usepackage{mathbbol}
\DeclareSymbolFontAlphabet{\mathbb}{AMSb} % AMS bold is prettier ...
\DeclareSymbolFontAlphabet{\mathbbl}{bbold} % but needs more symbols.

\newcommand{\ej}{\ensuremath{E_{\mathrm{J}}}}
\newcommand{\ec}{\ensuremath{E_{\mathrm{c}}}}

\newcommand{\cre}[1]{\hat{#1}^\dagger}
\newcommand{\anh}[1]{\hat{#1}^{\phantom{\dagger}}}

\newcommand{\ket}[1]{\ensuremath{|#1\rangle}\xspace}

\renewcommand{\Re}{\mathop{\mathrm{Re}}\nolimits}
\renewcommand{\mod}{\mathop{\mathrm{mod}}\nolimits}

\newcommand{\dorian}[1]{\textcolor{violet}{#1}}

\usepackage{amsmath}
\usepackage{color}

\usepackage{tikz, circuitikz}
\usetikzlibrary{shapes}
\usetikzlibrary{arrows}
\usetikzlibrary{math}
\usetikzlibrary{calc}
\usepackage{calc}

\begin{document}

\title{Direct detection of down-converted photons spontaneously produced at a single Josephson junction}

\author{Dorian Fraudet}	
\affiliation{Univ. Grenoble Alpes, CNRS, Grenoble INP, Institut Néel, 38000 Grenoble, France}
\author{Izak Snyman}
\affiliation{Mandelstam Institute for Theoretical Physics, School of Physics, University of the Witwatersrand,
Johannesburg, South Africa}
\author{Denis M. Basko}
\affiliation{Univ.~Grenoble Alpes, CNRS, LPMMC, 38000 Grenoble, France}
\author{Sébastien Léger}
\affiliation{Univ. Grenoble Alpes, CNRS, Grenoble INP, Institut Néel, 38000 Grenoble, France}
\affiliation{Department of Physics and Applied Physics, Stanford University, Stanford, California 94305, USA}
\author{Théo Sépulcre}
\affiliation{Wallenberg Centre for Quantum Technology, Chalmers University of Technology, SE-412 96 Gothenburg, Sweden}
\author{Arpit Ranadive}	
\affiliation{Univ. Grenoble Alpes, CNRS, Grenoble INP, Institut Néel, 38000 Grenoble, France}
\author{Gwenael Le Gal}	
\affiliation{Univ. Grenoble Alpes, CNRS, Grenoble INP, Institut Néel, 38000 Grenoble, France}
\author{Alba Torras-Coloma}
\affiliation{Institut de F\'isica d’Altes Energies (IFAE), The Barcelona Institute of
Science and Technology (BIST), Bellaterra (Barcelona) 08193, Spain}
\affiliation{Departament de F\'isica, Universitat Aut\`onoma de Barcelona, 08193 Bellaterra, Spain}
\author{Serge Florens}	
\affiliation{Univ. Grenoble Alpes, CNRS, Grenoble INP, Institut Néel, 38000 Grenoble, France}
\author{Nicolas Roch}	
\affiliation{Univ. Grenoble Alpes, CNRS, Grenoble INP, Institut Néel, 38000 Grenoble, France}

\begin{abstract}
We study spontaneous photon decay into multiple photons triggered by strong non-linearities in a superconducting quantum simulator of the boundary sine-Gordon impurity model.
Previously, spectroscopic signatures of photon-conversion were reported and evidenced as resonances in the many-body spectrum of these systems.
Here, we report on the observation
of multi-mode fluorescence of a small Josephson junction embedded in a high impedance superconducting transmission line.
Measurement of the down-converted photons is achieved using state-of-the-art broadband parametric amplifiers. Photon triplet emission is 
explicitly demonstrated at a given frequency as the counterpart of inelastic photon decay at three-times the emission frequency. These results open exciting prospects for the burgeoning field of many-body quantum optics and offer a direct signature of the ultra-strong light-matter coupling.
\end{abstract}

\maketitle

\textit{Introduction}.---  Circuit Quantum Electrodynamics (cQED)~\cite{Blais.2021} constitutes an ideal playground for quantum simulation~\cite{Houck.2012, Carusotto.2020} due to the versatility in model Hamiltonian designs. The class of quantum many-body models whose simulation requires the least hardware overhead in cQED are quantum impurities.
They are well studied in traditional condensed matter, and can be obtained in cQED by coupling a large set of harmonic resonators to a single non-linear element placed at the boundary of the circuit. Such simulators offer tunability both for the ``speed of light'' propagating through the harmonic modes, and for the single ``atom'' properties, that are hard to engineer in the quantum optics context.
Reaching the true quantum many-body regime relies on two important pillars. First,
the super-strong coupling regime~\cite{Meiser.2006}, in which the impurity spectral width is larger than the free spectral range (the frequency spacing between two consecutive harmonic modes), is required to obtain broad-band photon emission. This 
was recently reported in cQED using very long distributed microwave resonators~\cite{Sundaresan.20155pg} or slow light (high impedance) transmission lines made of Josephson junction (JJ) arrays~\cite{Martinez.2019, Kuzmin.2019}. Second, strong non-linearity is required to open up the phase space of inelastic decay channels involving multi-photon states. This condition was demonstrated~\cite{FornDiaz2017,Leger.2019,Kuzmin.2021, 10.21468/scipostphys.14.5.130,Mehta.2023} for Josephson junctions designed with a small Josephson energy ($\ej$) to charging energy ($\ec$) ratio. Fully understanding experimental cQED quantum impurity simulators is a clear prerequisite before addressing the more challenging physics of bulk versions of the sine-Gordon model~\cite{Roy2021,Roy2023}.

Many-body extensions of several cornerstones in quantum optics have been demonstrated in recent years using cQED impurity simulators. For instance, the tiny Lamb shift of quantum optics was shown to turn into a giant renormalisation of the impurity frequency~\cite{FornDiaz2017,Leger.2019,10.21468/scipostphys.14.5.130}, an effect
originating from the ultra-strong coupling to a dissipative bath~\cite{Schon.1990,Hur.2012}.
In addition, extremely large inelastic photon losses were reported in Ref.~\cite{Kuzmin.2021, 10.21468/scipostphys.14.5.130}, and interpreted from multi-photon down-conversion processes, following theoretical predictions~\cite{Goldstein.2013,Sanchez-Burillo.2014, Gheeraert.2018,Houzet.2020, Burshtein.2021,burshtein2023inelastic,houzet2023microwave}. More direct evidence for multi-photon conversion was the remarkable observation of avoided crossings between multi-photon states~\cite{Mehta.2023}, which can be understood as a many-body version of the celebrated vacuum Rabi splitting~\cite{Mehta.2022tcj}. 
However, those studies relied mainly on single-frequency spectroscopic measurements, and
the direct evidence of photon down-conversion processes in quantum impurity simulators is still lacking.

In this work, we present the observation of down-converted photons produced in a quantum impurity system, highlighting the importance of frequency-conversion for dissipative structures in open quantum systems~\cite{Caldeira.1983}. Our setup can also be used as a source of traveling photon triplets, addressing a long-standing challenge in optics~\cite{Bencheikh.2007}.
Complementary to the stimulated up-conversion process observed by \textit{Vrajitoarea et al.}~\cite{Vrajitoarea.2022}, our demonstration verifies theoretical predictions~\cite{Kockum2017,Koshino.2022} that spontaneous frequency conversion is possible in the ultra-strong coupling regime without the need for a strong classical pump. This contrasts sharply with experiments on driven quantum impurities~\cite{Magazzu.2018} and with previous observations of frequency conversion relying on parametric processes~\cite{Zakka-Bajjani.2011pod, Svensson.2018kch, Chang.2020740z} or cascaded photon emission~\cite{Gasparinetti.2017}, where very strong drives were required.

\begin{figure}[htb]
\begin{center}
\includegraphics[width = 1.0\columnwidth]{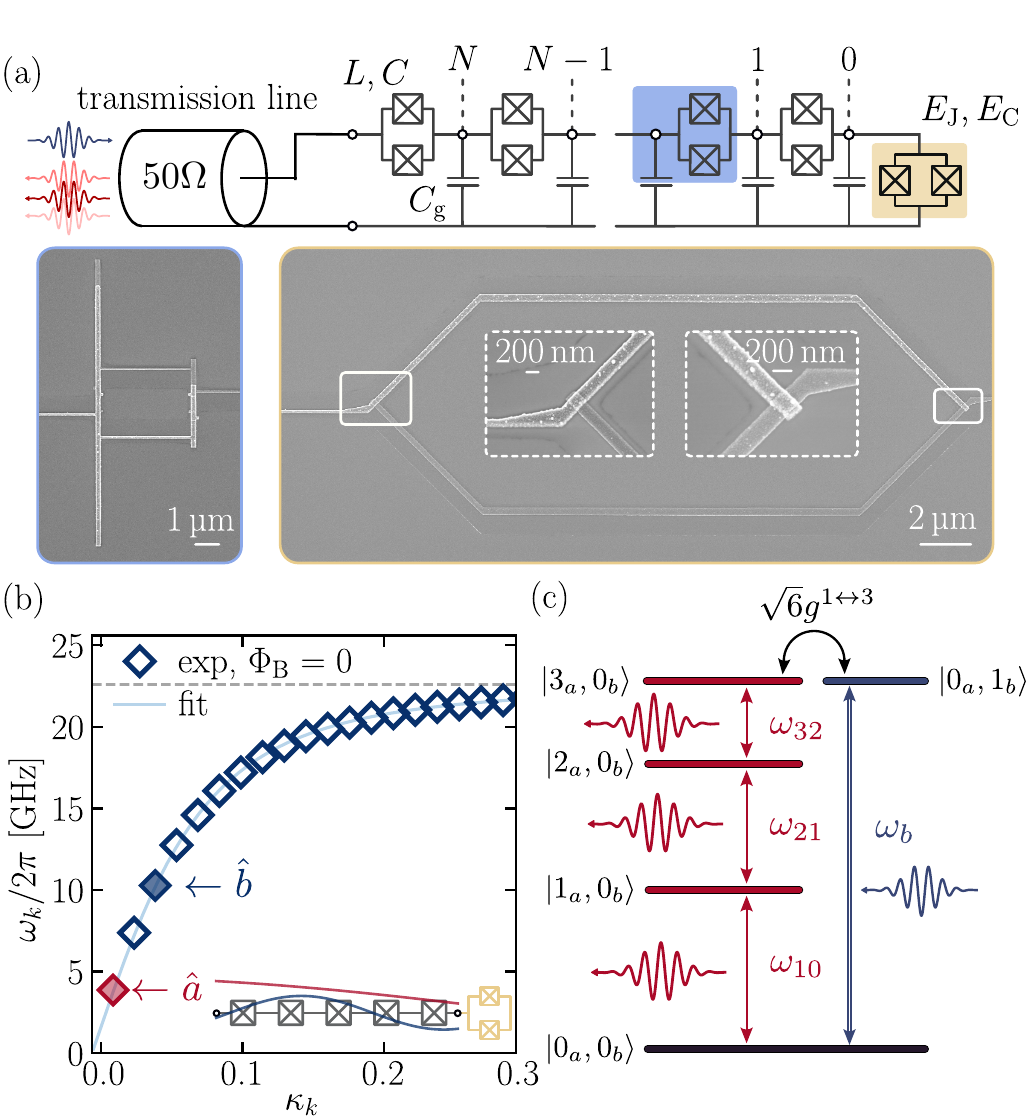}
\caption{{\bf (a)} Schematics of the measured circuit. The SQUID array (blue) is characterized by its flux-tunable lumped element inductance~$L$, capacitance $C$ and ground capacitance~$C_\text{g}$. The chain contains $N=200$ SQUIDs in the linear regime ($\displaystyle \ej/\ec=284$ at zero flux) and is terminated by a nonlinear SQUID ($\displaystyle \ej/\ec\sim 3$ at zero flux), depicted in yellow and characterized by the flux-tunable Josephson energy $E_\text{J}(\Phi_\text{B})$ and capacitance $C_{\rm J}$.
 SEM pictures: in the blue contour, image of a single SQUID of the array corresponding to the blue shaded area of panel~(a); in the yellow contour, image of the nonlinear SQUID corresponding to the gold shaded area of panel (a).
\textbf{(b)} Measured dispersion relation of the device showing the resonance frequencies $\omega_k$  of the first 19 modes as a function of their dimensionless wave-vector $\kappa_k$ ($0<\kappa<\pi$). Relevant modes to the experiment are highlighted in red (mode~$\hat{a}$) and blue (mode $\displaystyle \hat{b}$). Inset shows a sketch of their spatial profiles $\phi_{i,a}$ (red) and $\phi_{i,b}$ (blue) on sites $i=0,\ldots,N$. The measured dispersion relation is fitted (plain blue line) to determine $\displaystyle C_g$ (see Sec.~A in Supplemental Material). \textbf{(c)}  Energy levels and transitions involved in the three-photon down-conversion fluorescence. 
}
\label{fig1}
\end{center}
\end{figure}

{\it Experimental setup.--- }
In previous quantum impurity simulator experiments~\cite{Leger.2019,Kuzmin.2021, 10.21468/scipostphys.14.5.130}, it was extremely challenging to observe the down-converted photons because there were too many possible final states for the decay of a photon with frequency $\omega_\textrm{in}$ into several photons with frequencies $\omega_k$ fulfilling $\omega_\textrm{in}=\Sigma_{k} \omega_k$.  The output power at each mode $\omega_k$ was simply too low to be resolved within a reasonable integration time. To circumvent this problem, we designed here a shorter transmission line, leading to a larger free spectral range of the order of $3\ \textrm{GHz}$. Both the quantum impurity and the transmission line are made of SQUIDs, allowing us to fine-tune the frequency of the modes in-situ and reach the resonant condition $\omega_\textrm{in}=3\omega_\textrm{out}$ favouring the generation of photon triplets at frequency $\omega_\textrm{out}$. The device under study, depicted in Figure \ref{fig1}, consists of a microwave multimode Fabry-Perot resonator represented by an array of $N=200$ JJs. The array junctions are in the linear regime ($\ej^{\mathrm{arr}}/\ec^{\mathrm{arr}} \sim 300$) and act as high inductances $L=\left(\hbar / 2 e\right)^2 / \ej^{\mathrm{arr}}$. The JJ array thus implements a high-impedance transmission line characterized by its impedance $Z_\text{A} = \sqrt{L/C_g} < R_Q$ with the superconducting resistance quantum $R_Q \equiv h/(2e)^2 \simeq 6.5~\text{k}\Omega$. This transmission line is turned into a multi-mode microwave resonator by terminating it with impedance-unmatched loads at both ends. The left boundary (site $i=N+1$) is connected to a $\mathrm{50}\:\Omega$ measurement line and the right boundary (site $i=0$) is galvanically coupled to a small Josephson junction ($\ej/\ec \sim 3$) acting as a nonlinear boundary condition. This small junction is characterized by its superconducting phase difference $\hat{\phi}_0$. The small $\ej/\ec$ ratio together with the large array impedance  $Z_\text{A}\sim 0.1 R_Q$ triggers strong phase fluctuations $\langle\hat{\phi}_0^2\rangle\simeq1$~\cite{Leger.2019}, making the terminal junction behave as an artificial atom.

The Hamiltonian describing the circuit of Fig.~\ref{fig1} is
\begin{equation}
\hat{H}=\sum_k \hbar \omega_k \hat{a}_k^{\dagger} \hat{a}_k-E_J\left(\cos \hat{\phi}_0-1+\hat{\phi}_0^2/2\right),\quad
\label{eq:fullham}
\end{equation}
where $\hat{\phi}_0=\sum_k \phi_{0,k}(\cre{a}_k+\anh{a}_k)$, the coefficients $\phi_{0,k}$ correspond to the amplitude of mode~$k$ at the small junction, while $\cre{a}_k$ and $\anh{a}_k$ are the normal mode creation and annihilation operators. Wavevectors $\kappa_k$, resonance frequencies $\omega_k$ and spatial profiles $\phi_{i,k}$ of the normal modes are determined by solving the generalized eigenvalue problem of the linearized circuit. Figure~\ref{fig1}(b) shows the measured frequency of the first 19 modes (out of 200) of the system as a function of their dimensionless wavevectors $\kappa$. The second term in Eq.~(\ref{eq:fullham}) is the non-linear part of the small junction cosine potential, $\cos{\hat{\phi}_0} - 1 + \hat{\phi}_0^2/2 = \sum_{n>1} (-1)^{n}\hat{\phi}_0^{2n}/[(2n)!]$, the quadratic part already included in the normal mode decomposition.

This non-linearity induces interactions between the array modes. Indeed, substituting $\hat{\phi}_0=\sum_k \phi_{0,k}(\cre{a}_k+\anh{a}_k)$ in the non-linear term of the Hamiltonian, one readily obtains many terms of the form $\cre{a}_{k_n} \ldots \cre{a}_{k_1} \anh{a}_{l_m}\ldots\anh{a}_{l_1}$ that do not conserve the number of photons (with an even total number of operators). In particular, the nonlinear part of Eq.~(\ref{eq:fullham}) contains a term of the form $\cre{a}_{k_1}\cre{a}_{k_2}\cre{a}_{k_3}\anh{a}_{k_4}$, corresponding to splitting a single photon in mode $k_4$ into three photons in modes $k_1$, $k_2$ and $k_3$. In our experiment, the sample is designed such that the conversion process happens between two modes, \textit{i.~e.} $k_1 = k_2 = k_3$, and a single photon of mode $k_4$ is converted into three photons of mode $k_1$. In the rest of the paper, we denote the two relevant modes by operators $\hat{a}$ (low-frequency mode) and $\hat{b}$ (high-frequency mode). The explicit form of the corresponding conversion term in Eq.~(\ref{eq:fullham}) is $\hat{H}_{ab}= \hbar g_{1\leftrightarrow 3}\left( \hat{a}^{\dagger}{}^3 \hat{b}+\hat{b}^{\dagger}\hat{a}^3 \right)$. If we take only the $\hat{\phi}_0^4$ term in the expansion of the cosine potential, the coupling strength is given by $\hbar g_{1\leftrightarrow 3}=(\ej/6) \phi_{0,a}^3 \phi_{0,b}$; higher-order terms can introduce corrections to this simple expression.

To achieve high conversion efficiency, it is essential to have strong phase zero-point fluctuations at the small junction site [see schematics in the inset of Fig.~\ref{fig1}(b)]. This requirement is met thanks to the high impedance of the JJ array. The second condition is that the single-photon state $|1_b\rangle$ of mode $\hat{b}$ and the three-photon state $|3_a\rangle$ of mode~$\hat{a}$ are resonant [Fig.~\ref{fig1}(c)]. The energy $E_{\left|3_a\right\rangle}$ of the latter state differs from $3\hbar\omega_a$ due to diagonal or self-Kerr terms of the nonlinear Hamiltonian. 
We designed the SQUIDs so that the loop area of the boundary SQUID with small $E_\text{J}$ is much larger than that of the array SQUIDs [see Fig.~\ref{fig1}(a)]. Then, a small variation of the external magnetic field significantly changes the flux~$\Phi_\text{B}$ through the boundary SQUID, and thus its Josephson energy~$E_\text{J}(\Phi_\text{B})$, without affecting much the flux through the array SQUIDs. This allows us to vary the detuning $E_{\left|3_a\right\rangle}-E_{\left|1_b\right\rangle}$. Large variations of the external magnetic field by multiples of flux quantum~$\Phi_0$ in the small junction loop leaves it unaffected, while changing sensibly the flux~$\Phi_\text{A}$ through the array SQUIDs. Such a variation allows us to study various resonance conditions between $|1_b\rangle$ and~$|3_a\rangle$, due to the modified array impedance, hence tuning different values of the resonance frequency.

\begin{figure}[htb]
	\begin{center}
		\includegraphics[width = 1.0\columnwidth]{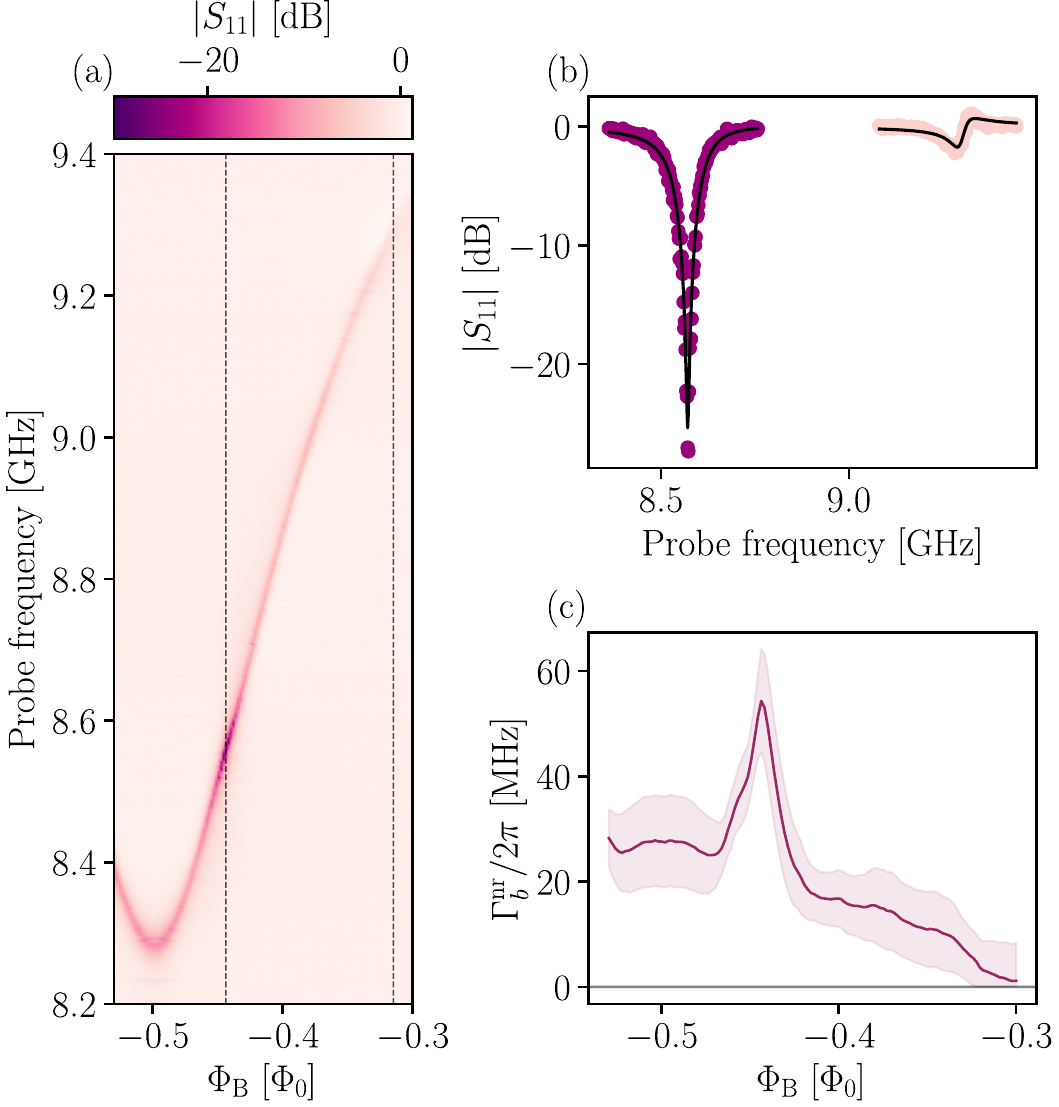}
		\caption{\textbf{(a)} Color map of the measured magnitude of the reflection coefficient $\left |S_{11}(\omega_\text{d}) \right |$ of mode $\displaystyle \hat{b}$ versus drive frequency $\omega_\text{d}$ and external magnetic flux $\Phi_\text{B}$ threading the small JJ loop. The mode frequency $\omega_b$ changes due to the flux dependence of the Josephson energy of the small JJ loop. A significant drop in $S_{11}$ around $\omega_b/2\pi = \qtylist{8.6}{\giga\hertz}$ and $\Phi_\text{B}=-0.44\,\Phi_0$ signals an enhancement of photon loss when $\omega_b \simeq 3 \omega_a$. Note that the fluorescence measurement was in fact carried out at a different arch. The corresponding data show a spurious mode that does not affect the analysis. They are available in Figure S1. 
        \textbf{(b)}~Slices corresponding to the vertical dashed lines of panel~(a) showing a drastic attenuation of the reflected signal close to $\Phi_\text{B}=-0.44\,\Phi_0$ attributed to photon down-conversion processes.
		\textbf{(c)} Non-radiative losses $\Gamma_b^{\mathrm{nr}}/2\pi$ of mode $\hat{b}$ as a function of $\Phi_\text{B}$ extracted by fitting $S_{11}(\omega_\text{d})$ to Eq.~(\ref{eq:reflin}). The shaded area around the data shows the uncertainty in the extracted non-radiative losses due to the imperfect isolation of the microwave components used in the output line~\cite{Rieger.20225gr}
        } 
		\label{fig2}
	\end{center}
\end{figure}

{\it Results.--- }
Prior to measuring the photon conversion process, we perform single-tone spectroscopy measurements to characterize our device and locate the external flux $\Phi_\text{B}$ leading to the multi-photon resonance condition. The sample is cooled down to a temperature $T\sim 12$~mK inside a dilution refrigerator. The reflection coefficient $S_{11}(\omega_\text{d})={a}_{\mathrm{out}}(\omega_\text{d})/{a}_{\mathrm{in}}(\omega_\text{d})$ is recorded using a Vector Network Analyzer, where ${a}_{\mathrm{in}}(\omega_\text{d})$ and ${a}_{\mathrm{out}}(\omega_\text{d})$ are respectively, the input and output fields at the drive frequency $\omega_\text{d}$. Fig.~\ref{fig2}(a) shows a map of the magnitude of the reflection coefficient $\left | S_{11}(\omega) \right|$ measured in a frequency window close to mode $b$ resonance frequency and for varying~$\Phi_\text{B}$.  From this measurement, we clearly see that $\left | S_{11}(\omega) \right |$ almost vanishes around $\Phi_\text{B} = -0.44 \Phi_0$.  Fig.~\ref{fig2}(b) shows constant flux cuts of $\left | S_{11}(\omega) \right |$ through the minimal value (purple) and far away from this minimum (pink). A drastic $-20\:\text{dB}$ attenuation of the reflected signal is observed. This decrease in the reflected signal amplitude implies an increase in losses of $\omega_b$ photons, that we attribute to the photon conversion process.

In the vicinity of mode $\hat{b}$ resonant frequency, the reflection coefficient is given by
\begin{equation}
S_{11}\left(\omega_{\mathrm{d}}\right)=\frac{\Gamma_b-2 \Gamma_b^{\mathrm{r}}+2i \left(\omega_b-\omega_\text{d}\right)}{\Gamma_b \dorian{-}2i \left(\omega_b-\omega_\text{d}\right)} \, ,
\label{eq:reflin}
\end{equation}
where $\Gamma_b$ is the total photon decay rate in mode~$b$, while  $\Gamma_b^{\mathrm{r}}$ is the radiative decay rate due to coupling to the measurement line. Figure \ref{fig2}(c) shows the fitted values of non-radiative decay rate $\Gamma_b^{\mathrm{nr}}\equiv\Gamma_b-\Gamma_b^{\mathrm{r}}$ as a function of the external flux $\Phi_\text{B}$. From this measurement, we clearly see a sharp increase in the internal losses of mode $b$, going from a baseline $\Gamma_b^{\mathrm{nr,0}} \simeq 2\pi\times 20$ MHz to $\Gamma_b^{\mathrm{nr}} = \Gamma_b^{\mathrm{nr,0}} + \Gamma^{1\leftrightarrow 3 } = 2\pi\times60$ MHz on resonance. 
We repeated the experiment at other values of the external flux, leading to lower array impedances, and to a resonance condition $\omega_a\simeq 3 \omega_b$ occurring at different frequencies. At the lowest impedance, we found a seven-fold increase of $\Gamma_b^{\mathrm{nr}}$. Thus, the observed loss enhancement is not due to a spurious resonance which would be expected to have a fixed frequency (see Sec.~B in Supplemental Material).
At this flux bias point, the three-photon conversion losses overcome other non-radiative losses, namely $\Gamma^{1\leftrightarrow3}>\Gamma_b^{\mathrm{nr,0}}$. Such inelastic effects induced by a boundary were already reported~\cite{Kuzmin.2021, 10.21468/scipostphys.14.5.130}. We will demonstrate now that their microscopic origin is indeed photon down-conversion.

\begin{figure}[htb]
	\begin{center}
		\includegraphics[width = 1.0\columnwidth]{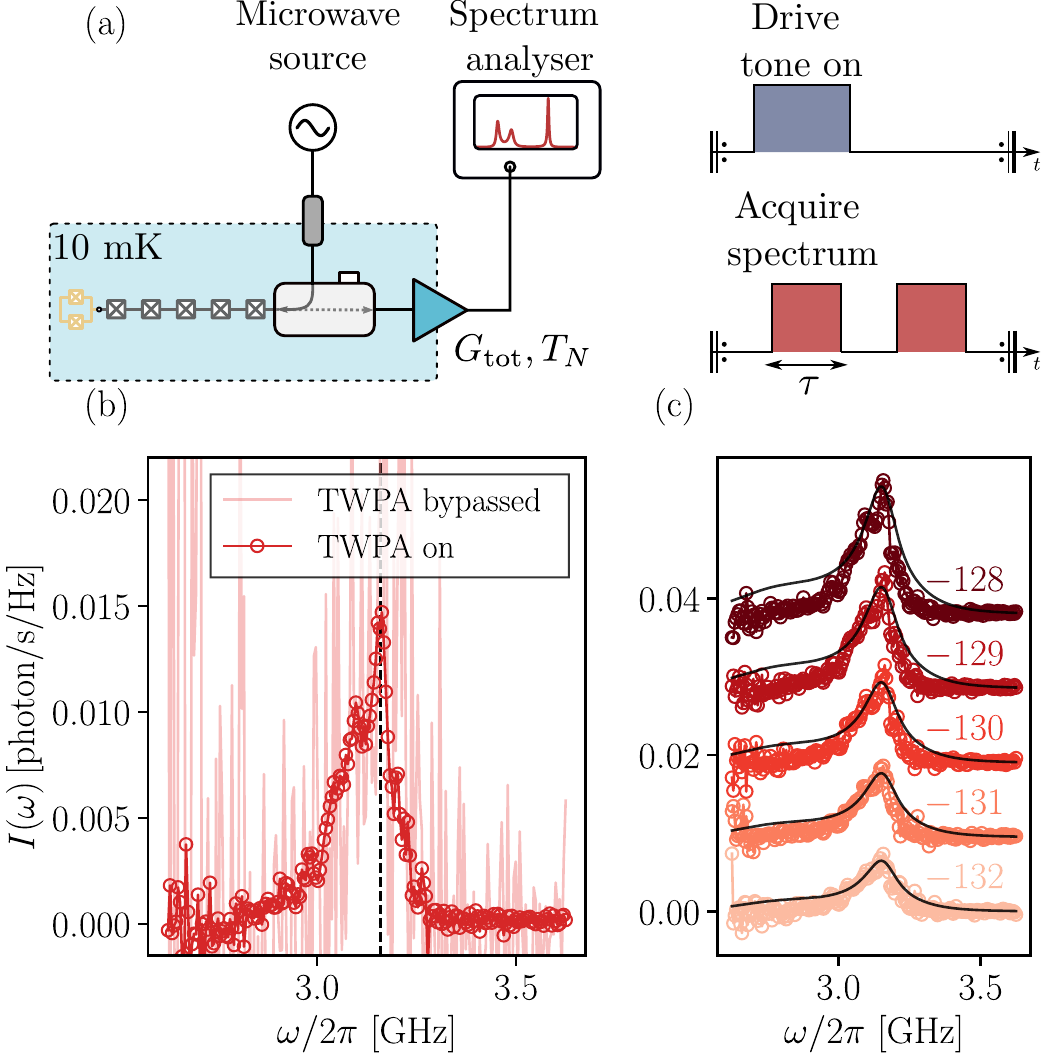}
		\caption{Spectrum of the emitted down-converted photons.
		 \textbf{(a)} Left panel: A simplified schematic of the measurement setup. Right panel: On-off measurement scheme.
		 \textbf{(b)} Emission spectrum at a fixed drive frequency $\omega_\textrm{d} = \omega_b = 2\pi\times9.13\:\text{GHz}$ with (dark red) and without (light red) TWPA for a drive power $P_\mathrm{in}= -129\:\text{dBm}$. 
		  \textbf{(c)} Emission spectra for increasing driving tone power $P_\mathrm{in}$ (from light red to dark red, vertically shifted by 0.01 for clarity). The values of $P_\mathrm{in}$ are given in dBm by the colored negative numbers. Black solid lines show the emission spectra calculated theoretically. All shown data were acquired at $\Phi_\textrm{B}=-0.44\,\Phi_0$.}
		\label{fig3}
	\end{center}
\end{figure}
Our measurement starts by setting the external flux to the value $\Phi_\text{B}^{\mathrm{opt}}$ corresponding to the peak in the inelastic losses $\Gamma_b^{\mathrm{nr}}$ [Fig.~\ref{fig2}(b)]. The system is then continuously driven at frequency $\omega_b/2\pi=9.13$ GHz. Anticipating triplet emission, the spectra are acquired on a 1 GHz span around $\omega_a/2\pi \simeq 3.16$ GHz with a resolution bandwidth of 5~MHz using a Power Spectrum Analyser, with a sweep time of 1 s. To circumvent low frequency drifts that would completely blur the measurement, the spectrum is alternatively measured with driving tone and without driving tone [Fig.~\ref{fig3}(a)]. Each on-off sequence is repeated 22500 times. Figure \ref{fig3}(b) shows the spectrum measured with (dark red curve) and without (light red curve) a Traveling Wave Parametric Amplifier TWPA)~\cite{Ranadive.2022} for otherwise identical experimental parameters. Considering the total measurement time of about 20 hours required to obtain a signal with very low signal to noise ratio (SNR) using only a low-noise amplifier  as the first stage of amplification, this measurement clearly demonstrates the advantages of using a TWPA in such a low-power microwave quantum optics experiment. More specifically, the TWPA can improve the SNR up to 14~dB. Interestingly, this improved SNR allows us to resolve the asymmetric shape of the fluorescence peak, which can be attributed to the anharmonicity of mode $\hat{a}$ as will be discussed later.
We have checked that the fluorescence signal at $\omega_a\simeq\omega_b/3$ quickly disappears when detuning the pump from frequency $\omega_b$, showing that the low frequency photons are indeed coming from down-conversion. 
Another important feature of the fluorescence signal is its dependence on the drive power $P_\mathrm{in}$. We measured several emission spectra for different values of $P_\mathrm{in}$, while still remaining in the single-photon regime. The resulting spectra are shown in Figure~\ref{fig3}~(c). The output power increases linearly with input power $P_\mathrm{in}$, as expected for a spontaneous down-conversion process  (see Sec.~C in Supplemental Material).

{\it Theoretical model.---}
To understand the asymmetric spectral shape of the measured fluorescence signal, we calculate it theoretically using the quantum regression theorem for a multi-level system coupled to a radiation field~\cite{Mollow.1969}. (See Sec.~D of the Supplemental Material for more detail.) Focusing on the weak drive limit, we truncate the system Hilbert space to the five levels shown in  Fig.~\ref{fig1}(c): $\ket{0_a0_b},\ket{0_a1_b},\ket{3_a0_b},\ket{2_a0_b},\ket{1_a0_b}$ with energies $0,\hbar\omega_b,\hbar(\omega_{32} + \omega_{21} + \omega_{10}),\hbar(\omega_{21} + \omega_{10}),\hbar\omega_{10}$, respectively. Strictly speaking, $\omega_{10}$ and $\omega_b$ are different from the harmonic frequencies $\omega_k$ in Eq.~(\ref{eq:fullham}) due to nonlinear corrections; however, they are determined experimentally by the reflection measurement ($2\pi\times3.16\:\mbox{GHz}$ and $2\pi\times9.13\:\mbox{GHz}$, respectively), which also gives the modes radiative ($\Gamma^\text{r}_a=2\pi\times84\:\mbox{MHz}$, $\Gamma^\text{r}_b=2\pi\times53\:\mbox{MHz}$)
and non-radiative ($\Gamma_a^\text{nr}=2\pi\times57\,\text{MHz}$ and $\Gamma_b^\text{nr}=2\pi\times22\,\text{MHz}$) decay rates. 

The remaining parameters of the model are the higher transition frequencies $\omega_{32},\omega_{21}$ and the coupling $g_{1\leftrightarrow 3}$. In principle, they could be found from the $\hat\phi_0^4$ term of the cosine potential. 
Then, neglecting pure dephasing, we find the fluorescence spectrum consisting of three broadened peaks centered at $\omega_{10}$, $\omega_{21}$, and $\omega_\text{d}-\omega_{10}-\omega_{21}$ (indeed, if $\omega_\text{d}$ is detuned from $\omega_{32} + \omega_{21} + \omega_{10}$, the state $\ket{3_a0_b}$ is only virtually populated due to energy conservation, so there is no emission at frequency $\omega_{32}$).
This shape does not fully agree with the experiment, which leads us to conclude that $\hat\phi_0^4$ expansion of the cosine potential is not sufficient. Higher-order terms can produce rather complex nonlinear effects already at a few-photon level~\cite{Hriscu.2011, Smith.2023}, whose study is beyond the scope of the present paper. So we constrain $\omega_{32} + \omega_{21} + \omega_{10} = \omega_b$ (since the spectra were obtained at the flux corresponding to maximum conversion) and use $\omega_{21}$ and $g_{1\leftrightarrow 3}$ as global fitting parameters (the same values for each input power $P_\mathrm{in}$). The resulting values $\omega_{21}\approx2\pi\times 2.7\:\text{GHz}$ and $g_{1\leftrightarrow 3}\approx 2\pi\times7\:\text{MHz}$ give the theoretical curves in Fig.~\ref{fig3}(c), see Sec.~F of the Supplemental Material. We find reasonable agreement, despite the minimalistic nature of our model. Here, we have attributed all non-radiative losses to decay into an unmonitored reservoir,
which allows us to obtain a relatively compact theoretical expression for the spectral density of down-converted photons. (See Sec.~D of the Supplemental Material.)
We also investigated non-radiative broadening by pure dephasing in
Sec.~E and F in Supplemental Material. We find similar agreement when dephasing dominates.

{\it Conclusion and outlook.--- }
 We have observed three-photon down-conversion fluorescence for a quantum impurity simulator in circuit QED. This non-parametric process results in the spontaneous decay of a single photon into three photons of lower energy. 
First, a microwave reflection experiment revealed a strong resonant-like depletion of the drive tone suggesting the presence of an additional loss channel related to the resonant photon conversion process. Second, using state-of-the-art near-quantum-limited parametric amplification, we could directly measure the down-converted photons, which constitutes, to the best of our knowledge, the first model-independent detection of a predicted ultra-strong coupling effect~\cite{Kockum2017,Koshino.2022}. This fluorescence spectrum is well reproduced by an effective few-level  model.
Our finding shows that microwave quantum optics techniques can shed new light on the physics of strongly interacting quantum impurities. 

Our results also have implications for the fast-growing field of high-impedance superconducting quantum bits, which promise very long coherence times~\cite{Somoroff.2021, Gyenis.2021}. The non-linear couplings we have highlighted here form a non-negligible loss channel that could limit the coherence of these circuits, and understanding their potential impact will probably require models that go beyond the initial studies on this subject~\cite{Mizel.2019, Paolo.2019, 10.1103/physrevlett.127.237702}. This extremely efficient frequency conversion process could be an interesting alternative for generating Schrödinger cat states with an arbitrary number of components~\cite{Marquet.2023}.
Further investigation of the spontaneous down-conversion process could also involve the study of correlations between the emitted photons~\cite{Silva.2010}.
In addition, the future observation of the theoretically predicted~\cite{Goldstein.2013, Peropadre.2013, Sanchez-Burillo.2014, Gheeraert.2018,burshtein2023inelastic,houzet2023microwave}
broadband and universal inelastic down-conversion for long Josephson arrays involving many modes remains another important challenge for superconducting quantum simulators.

{\it Acknowledgments.--- }
This work was supported by the QuantERA grant SiUCs (Grant No.731473).
The sample was fabricated in the clean room facility
of Institute Neel, Grenoble. We sincerely thank all
the clean room staff for help with fabrication of the
devices. We would like to acknowledge E. Eyraud for
his extensive help in the installation and maintenance of
the cryogenic setup. We also thank J. Jarreau, D. Dufeu
and L. Del Rey for their support with the experimental
equipments. We are grateful to J. Ankerhold, M.
Resch and R. Albert for insightful discussions regarding this project and to Q. Ficheux et P. Forn-Diaz for their critical reading of the manuscript. We thank the members of the superconducting circuits
group at Neel Institute for helpful discussions.

\bibliography{biblio}

%apsrev4-2.bst 2019-01-14 (MD) hand-edited version of apsrev4-1.bst
%Control: key (0)
%Control: author (8) initials jnrlst
%Control: editor formatted (1) identically to author
%Control: production of article title (0) allowed
%Control: page (0) single
%Control: year (1) truncated
%Control: production of eprint (0) enabled
\begin{thebibliography}{48}%
\makeatletter
\providecommand \@ifxundefined [1]{%
 \@ifx{#1\undefined}
}%
\providecommand \@ifnum [1]{%
 \ifnum #1\expandafter \@firstoftwo
 \else \expandafter \@secondoftwo
 \fi
}%
\providecommand \@ifx [1]{%
 \ifx #1\expandafter \@firstoftwo
 \else \expandafter \@secondoftwo
 \fi
}%
\providecommand \natexlab [1]{#1}%
\providecommand \enquote  [1]{``#1''}%
\providecommand \bibnamefont  [1]{#1}%
\providecommand \bibfnamefont [1]{#1}%
\providecommand \citenamefont [1]{#1}%
\providecommand \href@noop [0]{\@secondoftwo}%
\providecommand \href [0]{\begingroup \@sanitize@url \@href}%
\providecommand \@href[1]{\@@startlink{#1}\@@href}%
\providecommand \@@href[1]{\endgroup#1\@@endlink}%
\providecommand \@sanitize@url [0]{\catcode `\\12\catcode `\$12\catcode `\&12\catcode `\#12\catcode `\^12\catcode `\_12\catcode `\%12\relax}%
\providecommand \@@startlink[1]{}%
\providecommand \@@endlink[0]{}%
\providecommand \url  [0]{\begingroup\@sanitize@url \@url }%
\providecommand \@url [1]{\endgroup\@href {#1}{\urlprefix }}%
\providecommand \urlprefix  [0]{URL }%
\providecommand \Eprint [0]{\href }%
\providecommand \doibase [0]{https://doi.org/}%
\providecommand \selectlanguage [0]{\@gobble}%
\providecommand \bibinfo  [0]{\@secondoftwo}%
\providecommand \bibfield  [0]{\@secondoftwo}%
\providecommand \translation [1]{[#1]}%
\providecommand \BibitemOpen [0]{}%
\providecommand \bibitemStop [0]{}%
\providecommand \bibitemNoStop [0]{.\EOS\space}%
\providecommand \EOS [0]{\spacefactor3000\relax}%
\providecommand \BibitemShut  [1]{\csname bibitem#1\endcsname}%
\let\auto@bib@innerbib\@empty
%</preamble>
\bibitem [{\citenamefont {Blais}\ \emph {et~al.}(2021)\citenamefont {Blais}, \citenamefont {Grimsmo}, \citenamefont {Girvin},\ and\ \citenamefont {Wallraff}}]{Blais.2021}%
  \BibitemOpen
  \bibfield  {author} {\bibinfo {author} {\bibfnamefont {A.}~\bibnamefont {Blais}}, \bibinfo {author} {\bibfnamefont {A.~L.}\ \bibnamefont {Grimsmo}}, \bibinfo {author} {\bibfnamefont {S.~M.}\ \bibnamefont {Girvin}},\ and\ \bibinfo {author} {\bibfnamefont {A.}~\bibnamefont {Wallraff}},\ }\bibfield  {title} {\bibinfo {title} {{Circuit quantum electrodynamics}},\ }\href {https://doi.org/10.1103/revmodphys.93.025005} {\bibfield  {journal} {\bibinfo  {journal} {Reviews of Modern Physics}\ }\textbf {\bibinfo {volume} {93}},\ \bibinfo {pages} {025005} (\bibinfo {year} {2021})}\BibitemShut {NoStop}%
\bibitem [{\citenamefont {Houck}\ \emph {et~al.}(2012)\citenamefont {Houck}, \citenamefont {T{\"u}reci},\ and\ \citenamefont {Koch}}]{Houck.2012}%
  \BibitemOpen
  \bibfield  {author} {\bibinfo {author} {\bibfnamefont {A.~A.}\ \bibnamefont {Houck}}, \bibinfo {author} {\bibfnamefont {H.~E.}\ \bibnamefont {T{\"u}reci}},\ and\ \bibinfo {author} {\bibfnamefont {J.}~\bibnamefont {Koch}},\ }\bibfield  {title} {\bibinfo {title} {{On-chip quantum simulation with superconducting circuits}},\ }\href {https://doi.org/10.1038/nphys2251} {\bibfield  {journal} {\bibinfo  {journal} {Nature Physics}\ }\textbf {\bibinfo {volume} {8}},\ \bibinfo {pages} {292 } (\bibinfo {year} {2012})}\BibitemShut {NoStop}%
\bibitem [{\citenamefont {Carusotto}\ \emph {et~al.}(2020)\citenamefont {Carusotto}, \citenamefont {Houck}, \citenamefont {Koll{\'a}r}, \citenamefont {Roushan}, \citenamefont {Schuster},\ and\ \citenamefont {Simon}}]{Carusotto.2020}%
  \BibitemOpen
  \bibfield  {author} {\bibinfo {author} {\bibfnamefont {I.}~\bibnamefont {Carusotto}}, \bibinfo {author} {\bibfnamefont {A.~A.}\ \bibnamefont {Houck}}, \bibinfo {author} {\bibfnamefont {A.~J.}\ \bibnamefont {Koll{\'a}r}}, \bibinfo {author} {\bibfnamefont {P.}~\bibnamefont {Roushan}}, \bibinfo {author} {\bibfnamefont {D.~I.}\ \bibnamefont {Schuster}},\ and\ \bibinfo {author} {\bibfnamefont {J.}~\bibnamefont {Simon}},\ }\bibfield  {title} {\bibinfo {title} {Photonic materials in circuit quantum electrodynamics},\ }\href {https://doi.org/10.1038/s41567-020-0815-y} {\bibfield  {journal} {\bibinfo  {journal} {Nature Physics}\ }\textbf {\bibinfo {volume} {16}},\ \bibinfo {pages} {268} (\bibinfo {year} {2020})}\BibitemShut {NoStop}%
\bibitem [{\citenamefont {Meiser}\ and\ \citenamefont {Meystre}(2006)}]{Meiser.2006}%
  \BibitemOpen
  \bibfield  {author} {\bibinfo {author} {\bibfnamefont {D.}~\bibnamefont {Meiser}}\ and\ \bibinfo {author} {\bibfnamefont {P.}~\bibnamefont {Meystre}},\ }\bibfield  {title} {\bibinfo {title} {{Superstrong coupling regime of cavity quantum electrodynamics}},\ }\href {https://doi.org/10.1103/physreva.74.065801} {\bibfield  {journal} {\bibinfo  {journal} {Physical Review A}\ }\textbf {\bibinfo {volume} {74}},\ \bibinfo {pages} {065801 } (\bibinfo {year} {2006})}\BibitemShut {NoStop}%
\bibitem [{\citenamefont {Sundaresan}\ \emph {et~al.}(2015)\citenamefont {Sundaresan}, \citenamefont {Liu}, \citenamefont {Sadri}, \citenamefont {Szocs}, \citenamefont {Underwood}, \citenamefont {Malekakhlagh}, \citenamefont {T{\"u}reci},\ and\ \citenamefont {Houck}}]{Sundaresan.20155pg}%
  \BibitemOpen
  \bibfield  {author} {\bibinfo {author} {\bibfnamefont {N.~M.}\ \bibnamefont {Sundaresan}}, \bibinfo {author} {\bibfnamefont {Y.}~\bibnamefont {Liu}}, \bibinfo {author} {\bibfnamefont {D.}~\bibnamefont {Sadri}}, \bibinfo {author} {\bibfnamefont {L.~J.}\ \bibnamefont {Szocs}}, \bibinfo {author} {\bibfnamefont {D.~L.}\ \bibnamefont {Underwood}}, \bibinfo {author} {\bibfnamefont {M.}~\bibnamefont {Malekakhlagh}}, \bibinfo {author} {\bibfnamefont {H.~E.}\ \bibnamefont {T{\"u}reci}},\ and\ \bibinfo {author} {\bibfnamefont {A.~A.}\ \bibnamefont {Houck}},\ }\bibfield  {title} {\bibinfo {title} {{Beyond Strong Coupling in a Multimode Cavity}},\ }\href {https://doi.org/10.1103/physrevx.5.021035} {\bibfield  {journal} {\bibinfo  {journal} {Physical Review X}\ }\textbf {\bibinfo {volume} {5}},\ \bibinfo {pages} {021035 } (\bibinfo {year} {2015})}\BibitemShut {NoStop}%
\bibitem [{\citenamefont {Martinez}\ \emph {et~al.}(2019)\citenamefont {Martinez}, \citenamefont {Leger}, \citenamefont {Gheeraert}, \citenamefont {Dassonneville}, \citenamefont {Planat}, \citenamefont {Foroughi}, \citenamefont {Krupko}, \citenamefont {Buisson}, \citenamefont {Naud}, \citenamefont {Hasch-Guichard}, \citenamefont {Florens}, \citenamefont {Snyman},\ and\ \citenamefont {Roch}}]{Martinez.2019}%
  \BibitemOpen
  \bibfield  {author} {\bibinfo {author} {\bibfnamefont {J.~P.}\ \bibnamefont {Martinez}}, \bibinfo {author} {\bibfnamefont {S.}~\bibnamefont {Leger}}, \bibinfo {author} {\bibfnamefont {N.}~\bibnamefont {Gheeraert}}, \bibinfo {author} {\bibfnamefont {R.}~\bibnamefont {Dassonneville}}, \bibinfo {author} {\bibfnamefont {L.}~\bibnamefont {Planat}}, \bibinfo {author} {\bibfnamefont {F.}~\bibnamefont {Foroughi}}, \bibinfo {author} {\bibfnamefont {Y.}~\bibnamefont {Krupko}}, \bibinfo {author} {\bibfnamefont {O.}~\bibnamefont {Buisson}}, \bibinfo {author} {\bibfnamefont {C.}~\bibnamefont {Naud}}, \bibinfo {author} {\bibfnamefont {W.}~\bibnamefont {Hasch-Guichard}}, \bibinfo {author} {\bibfnamefont {S.}~\bibnamefont {Florens}}, \bibinfo {author} {\bibfnamefont {I.}~\bibnamefont {Snyman}},\ and\ \bibinfo {author} {\bibfnamefont {N.}~\bibnamefont {Roch}},\ }\bibfield  {title} {\bibinfo {title} {{A tunable Josephson platform to explore many-body quantum optics in circuit-QED}},\ }\href
  {https://doi.org/10.1038/s41534-018-0104-0} {\bibfield  {journal} {\bibinfo  {journal} {npj Quantum Information}\ }\textbf {\bibinfo {volume} {5}},\ \bibinfo {pages} {1829} (\bibinfo {year} {2019})}\BibitemShut {NoStop}%
\bibitem [{\citenamefont {Kuzmin}\ \emph {et~al.}(2019)\citenamefont {Kuzmin}, \citenamefont {Mehta}, \citenamefont {Grabon}, \citenamefont {Mencia},\ and\ \citenamefont {Manucharyan}}]{Kuzmin.2019}%
  \BibitemOpen
  \bibfield  {author} {\bibinfo {author} {\bibfnamefont {R.}~\bibnamefont {Kuzmin}}, \bibinfo {author} {\bibfnamefont {N.}~\bibnamefont {Mehta}}, \bibinfo {author} {\bibfnamefont {N.}~\bibnamefont {Grabon}}, \bibinfo {author} {\bibfnamefont {R.}~\bibnamefont {Mencia}},\ and\ \bibinfo {author} {\bibfnamefont {V.~E.}\ \bibnamefont {Manucharyan}},\ }\bibfield  {title} {\bibinfo {title} {{Superstrong coupling in circuit quantum electrodynamics}},\ }\href {https://doi.org/10.1038/s41534-019-0134-2} {\bibfield  {journal} {\bibinfo  {journal} {npj Quantum Information}\ }\textbf {\bibinfo {volume} {5}},\ \bibinfo {pages} {1 } (\bibinfo {year} {2019})}\BibitemShut {NoStop}%
\bibitem [{\citenamefont {Forn-Diaz}\ \emph {et~al.}(2017)\citenamefont {Forn-Diaz}, \citenamefont {Garcia-Ripoll}, \citenamefont {Peropadre}, \citenamefont {Orgiazzi}, \citenamefont {Yurtalan}, \citenamefont {Belyansky}, \citenamefont {Wilson},\ and\ \citenamefont {Lupascu}}]{FornDiaz2017}%
  \BibitemOpen
  \bibfield  {author} {\bibinfo {author} {\bibfnamefont {P.}~\bibnamefont {Forn-Diaz}}, \bibinfo {author} {\bibfnamefont {J.~J.}\ \bibnamefont {Garcia-Ripoll}}, \bibinfo {author} {\bibfnamefont {B.}~\bibnamefont {Peropadre}}, \bibinfo {author} {\bibfnamefont {J.-L.}\ \bibnamefont {Orgiazzi}}, \bibinfo {author} {\bibfnamefont {M.~A.}\ \bibnamefont {Yurtalan}}, \bibinfo {author} {\bibfnamefont {R.}~\bibnamefont {Belyansky}}, \bibinfo {author} {\bibfnamefont {C.~M.}\ \bibnamefont {Wilson}},\ and\ \bibinfo {author} {\bibfnamefont {A.}~\bibnamefont {Lupascu}},\ }\bibfield  {title} {\bibinfo {title} {{Ultrastrong coupling of a single artificial atom to an electromagnetic continuum in the nonperturbative regime}},\ }\href {https://doi.org/10.1038/nphys3905} {\bibfield  {journal} {\bibinfo  {journal} {Nature Physics}\ }\textbf {\bibinfo {volume} {13}},\ \bibinfo {pages} {39} (\bibinfo {year} {2017})}\BibitemShut {NoStop}%
\bibitem [{\citenamefont {L{\'e}ger}\ \emph {et~al.}(2019)\citenamefont {L{\'e}ger}, \citenamefont {Puertas-Mart{\'i}nez}, \citenamefont {Bharadwaj}, \citenamefont {Dassonneville}, \citenamefont {Delaforce}, \citenamefont {Foroughi}, \citenamefont {Milchakov}, \citenamefont {Planat}, \citenamefont {Buisson}, \citenamefont {Naud}, \citenamefont {Hasch-Guichard}, \citenamefont {Florens}, \citenamefont {Snyman},\ and\ \citenamefont {Roch}}]{Leger.2019}%
  \BibitemOpen
  \bibfield  {author} {\bibinfo {author} {\bibfnamefont {S.}~\bibnamefont {L{\'e}ger}}, \bibinfo {author} {\bibfnamefont {J.}~\bibnamefont {Puertas-Mart{\'i}nez}}, \bibinfo {author} {\bibfnamefont {K.}~\bibnamefont {Bharadwaj}}, \bibinfo {author} {\bibfnamefont {R.}~\bibnamefont {Dassonneville}}, \bibinfo {author} {\bibfnamefont {J.}~\bibnamefont {Delaforce}}, \bibinfo {author} {\bibfnamefont {F.}~\bibnamefont {Foroughi}}, \bibinfo {author} {\bibfnamefont {V.}~\bibnamefont {Milchakov}}, \bibinfo {author} {\bibfnamefont {L.}~\bibnamefont {Planat}}, \bibinfo {author} {\bibfnamefont {O.}~\bibnamefont {Buisson}}, \bibinfo {author} {\bibfnamefont {C.}~\bibnamefont {Naud}}, \bibinfo {author} {\bibfnamefont {W.}~\bibnamefont {Hasch-Guichard}}, \bibinfo {author} {\bibfnamefont {S.}~\bibnamefont {Florens}}, \bibinfo {author} {\bibfnamefont {I.}~\bibnamefont {Snyman}},\ and\ \bibinfo {author} {\bibfnamefont {N.}~\bibnamefont {Roch}},\ }\bibfield  {title} {\bibinfo {title} {Observation of quantum many-body effects due
  to zero point fluctuations in superconducting circuits},\ }\href {https://doi.org/10.1038/s41467-019-13199-x} {\bibfield  {journal} {\bibinfo  {journal} {Nature Communications}\ }\textbf {\bibinfo {volume} {10}},\ \bibinfo {pages} {5259} (\bibinfo {year} {2019})}\BibitemShut {NoStop}%
\bibitem [{\citenamefont {Kuzmin}\ \emph {et~al.}(2021)\citenamefont {Kuzmin}, \citenamefont {Grabon}, \citenamefont {Mehta}, \citenamefont {Burshtein}, \citenamefont {Goldstein}, \citenamefont {Houzet}, \citenamefont {Glazman},\ and\ \citenamefont {Manucharyan}}]{Kuzmin.2021}%
  \BibitemOpen
  \bibfield  {author} {\bibinfo {author} {\bibfnamefont {R.}~\bibnamefont {Kuzmin}}, \bibinfo {author} {\bibfnamefont {N.}~\bibnamefont {Grabon}}, \bibinfo {author} {\bibfnamefont {N.}~\bibnamefont {Mehta}}, \bibinfo {author} {\bibfnamefont {A.}~\bibnamefont {Burshtein}}, \bibinfo {author} {\bibfnamefont {M.}~\bibnamefont {Goldstein}}, \bibinfo {author} {\bibfnamefont {M.}~\bibnamefont {Houzet}}, \bibinfo {author} {\bibfnamefont {L.~I.}\ \bibnamefont {Glazman}},\ and\ \bibinfo {author} {\bibfnamefont {V.~E.}\ \bibnamefont {Manucharyan}},\ }\bibfield  {title} {\bibinfo {title} {{Inelastic Scattering of a Photon by a Quantum Phase Slip}},\ }\href {https://doi.org/10.1103/physrevlett.126.197701} {\bibfield  {journal} {\bibinfo  {journal} {Physical Review Letters}\ }\textbf {\bibinfo {volume} {126}},\ \bibinfo {pages} {197701} (\bibinfo {year} {2021})}\BibitemShut {NoStop}%
\bibitem [{\citenamefont {L{\'e}ger}\ \emph {et~al.}(2023)\citenamefont {L{\'e}ger}, \citenamefont {S{\'e}pulcre}, \citenamefont {Fraudet}, \citenamefont {Buisson}, \citenamefont {Naud}, \citenamefont {Hasch-Guichard}, \citenamefont {Florens}, \citenamefont {Snyman}, \citenamefont {Basko},\ and\ \citenamefont {Roch}}]{10.21468/scipostphys.14.5.130}%
  \BibitemOpen
  \bibfield  {author} {\bibinfo {author} {\bibfnamefont {S.}~\bibnamefont {L{\'e}ger}}, \bibinfo {author} {\bibfnamefont {T.}~\bibnamefont {S{\'e}pulcre}}, \bibinfo {author} {\bibfnamefont {D.}~\bibnamefont {Fraudet}}, \bibinfo {author} {\bibfnamefont {O.}~\bibnamefont {Buisson}}, \bibinfo {author} {\bibfnamefont {C.}~\bibnamefont {Naud}}, \bibinfo {author} {\bibfnamefont {W.}~\bibnamefont {Hasch-Guichard}}, \bibinfo {author} {\bibfnamefont {S.}~\bibnamefont {Florens}}, \bibinfo {author} {\bibfnamefont {I.}~\bibnamefont {Snyman}}, \bibinfo {author} {\bibfnamefont {D.~M.}\ \bibnamefont {Basko}},\ and\ \bibinfo {author} {\bibfnamefont {N.}~\bibnamefont {Roch}},\ }\bibfield  {title} {\bibinfo {title} {{Revealing the finite-frequency response of a bosonic quantum impurity}},\ }\href {https://doi.org/10.21468/scipostphys.14.5.130} {\bibfield  {journal} {\bibinfo  {journal} {SciPost Physics}\ }\textbf {\bibinfo {volume} {14}},\ \bibinfo {pages} {130} (\bibinfo {year} {2023})}\BibitemShut {NoStop}%
\bibitem [{\citenamefont {Mehta}\ \emph {et~al.}(2023)\citenamefont {Mehta}, \citenamefont {Kuzmin}, \citenamefont {Ciuti},\ and\ \citenamefont {Manucharyan}}]{Mehta.2023}%
  \BibitemOpen
  \bibfield  {author} {\bibinfo {author} {\bibfnamefont {N.}~\bibnamefont {Mehta}}, \bibinfo {author} {\bibfnamefont {R.}~\bibnamefont {Kuzmin}}, \bibinfo {author} {\bibfnamefont {C.}~\bibnamefont {Ciuti}},\ and\ \bibinfo {author} {\bibfnamefont {V.~E.}\ \bibnamefont {Manucharyan}},\ }\bibfield  {title} {\bibinfo {title} {{Down-conversion of a single photon as a probe of many-body localization}},\ }\href {https://doi.org/10.1038/s41586-022-05615-y} {\bibfield  {journal} {\bibinfo  {journal} {Nature}\ }\textbf {\bibinfo {volume} {613}},\ \bibinfo {pages} {650} (\bibinfo {year} {2023})}\BibitemShut {NoStop}%
\bibitem [{\citenamefont {Roy}\ \emph {et~al.}(2021)\citenamefont {Roy}, \citenamefont {Schuricht}, \citenamefont {Hauschild}, \citenamefont {Pollmann},\ and\ \citenamefont {Saleur}}]{Roy2021}%
  \BibitemOpen
  \bibfield  {author} {\bibinfo {author} {\bibfnamefont {A.}~\bibnamefont {Roy}}, \bibinfo {author} {\bibfnamefont {D.}~\bibnamefont {Schuricht}}, \bibinfo {author} {\bibfnamefont {J.}~\bibnamefont {Hauschild}}, \bibinfo {author} {\bibfnamefont {F.}~\bibnamefont {Pollmann}},\ and\ \bibinfo {author} {\bibfnamefont {H.}~\bibnamefont {Saleur}},\ }\bibfield  {title} {\bibinfo {title} {The quantum sine-gordon model with quantum circuits},\ }\href {https://doi.org/https://doi.org/10.1016/j.nuclphysb.2021.115445} {\bibfield  {journal} {\bibinfo  {journal} {Nuclear Physics B}\ }\textbf {\bibinfo {volume} {968}},\ \bibinfo {pages} {115445} (\bibinfo {year} {2021})}\BibitemShut {NoStop}%
\bibitem [{\citenamefont {Roy}\ and\ \citenamefont {Lukyanov}(2023)}]{Roy2023}%
  \BibitemOpen
  \bibfield  {author} {\bibinfo {author} {\bibfnamefont {A.}~\bibnamefont {Roy}}\ and\ \bibinfo {author} {\bibfnamefont {S.~L.}\ \bibnamefont {Lukyanov}},\ }\bibfield  {title} {\bibinfo {title} {Soliton confinement in a quantum circuit},\ }\href {https://doi.org/10.1038/s41467-023-43107-3} {\bibfield  {journal} {\bibinfo  {journal} {Nature Communications}\ }\textbf {\bibinfo {volume} {14}},\ \bibinfo {pages} {7433} (\bibinfo {year} {2023})}\BibitemShut {NoStop}%
\bibitem [{\citenamefont {Sch\"on}\ and\ \citenamefont {Zaikin}(1990)}]{Schon.1990}%
  \BibitemOpen
  \bibfield  {author} {\bibinfo {author} {\bibfnamefont {G.}~\bibnamefont {Sch\"on}}\ and\ \bibinfo {author} {\bibfnamefont {A.~D.}\ \bibnamefont {Zaikin}},\ }\bibfield  {title} {\bibinfo {title} {{Quantum coherent effects, phase transitions, and the dissipative dynamics of ultra small tunnel junctions}},\ }\href {https://doi.org/10.1016/0370-1573(90)90156-v} {\bibfield  {journal} {\bibinfo  {journal} {Physics Reports}\ }\textbf {\bibinfo {volume} {198}},\ \bibinfo {pages} {237 } (\bibinfo {year} {1990})}\BibitemShut {NoStop}%
\bibitem [{\citenamefont {Le~Hur}(2012)}]{Hur.2012}%
  \BibitemOpen
  \bibfield  {author} {\bibinfo {author} {\bibfnamefont {K.}~\bibnamefont {Le~Hur}},\ }\bibfield  {title} {\bibinfo {title} {Kondo resonance of a microwave photon},\ }\href {https://doi.org/10.1103/PhysRevB.85.140506} {\bibfield  {journal} {\bibinfo  {journal} {Phys. Rev. B}\ }\textbf {\bibinfo {volume} {85}},\ \bibinfo {pages} {140506} (\bibinfo {year} {2012})}\BibitemShut {NoStop}%
\bibitem [{\citenamefont {Goldstein}\ \emph {et~al.}(2013)\citenamefont {Goldstein}, \citenamefont {Devoret}, \citenamefont {Houzet},\ and\ \citenamefont {Glazman}}]{Goldstein.2013}%
  \BibitemOpen
  \bibfield  {author} {\bibinfo {author} {\bibfnamefont {M.}~\bibnamefont {Goldstein}}, \bibinfo {author} {\bibfnamefont {M.~H.}\ \bibnamefont {Devoret}}, \bibinfo {author} {\bibfnamefont {M.}~\bibnamefont {Houzet}},\ and\ \bibinfo {author} {\bibfnamefont {L.~I.}\ \bibnamefont {Glazman}},\ }\bibfield  {title} {\bibinfo {title} {{Inelastic Microwave Photon Scattering off a Quantum Impurity in a Josephson-Junction Array}},\ }\href {https://doi.org/10.1103/physrevlett.110.017002} {\bibfield  {journal} {\bibinfo  {journal} {Physical Review Letters}\ }\textbf {\bibinfo {volume} {110}},\ \bibinfo {pages} {017002} (\bibinfo {year} {2013})}\BibitemShut {NoStop}%
\bibitem [{\citenamefont {Sanchez-Burillo}\ \emph {et~al.}(2014)\citenamefont {Sanchez-Burillo}, \citenamefont {Zueco}, \citenamefont {Garcia-Ripoll},\ and\ \citenamefont {Martin-Moreno}}]{Sanchez-Burillo.2014}%
  \BibitemOpen
  \bibfield  {author} {\bibinfo {author} {\bibfnamefont {E.}~\bibnamefont {Sanchez-Burillo}}, \bibinfo {author} {\bibfnamefont {D.}~\bibnamefont {Zueco}}, \bibinfo {author} {\bibfnamefont {J.~J.}\ \bibnamefont {Garcia-Ripoll}},\ and\ \bibinfo {author} {\bibfnamefont {L.}~\bibnamefont {Martin-Moreno}},\ }\bibfield  {title} {\bibinfo {title} {{Scattering in the Ultrastrong Regime: Nonlinear Optics with One Photon}},\ }\href {https://doi.org/10.1103/physrevlett.113.263604} {\bibfield  {journal} {\bibinfo  {journal} {Physical Review Letters}\ }\textbf {\bibinfo {volume} {113}},\ \bibinfo {pages} {263604 } (\bibinfo {year} {2014})}\BibitemShut {NoStop}%
\bibitem [{\citenamefont {Gheeraert}\ \emph {et~al.}(2018)\citenamefont {Gheeraert}, \citenamefont {Zhang}, \citenamefont {S{\'e}pulcre}, \citenamefont {Bera}, \citenamefont {Roch}, \citenamefont {Baranger},\ and\ \citenamefont {Florens}}]{Gheeraert.2018}%
  \BibitemOpen
  \bibfield  {author} {\bibinfo {author} {\bibfnamefont {N.}~\bibnamefont {Gheeraert}}, \bibinfo {author} {\bibfnamefont {X.~H.~H.}\ \bibnamefont {Zhang}}, \bibinfo {author} {\bibfnamefont {T.}~\bibnamefont {S{\'e}pulcre}}, \bibinfo {author} {\bibfnamefont {S.}~\bibnamefont {Bera}}, \bibinfo {author} {\bibfnamefont {N.}~\bibnamefont {Roch}}, \bibinfo {author} {\bibfnamefont {H.~U.}\ \bibnamefont {Baranger}},\ and\ \bibinfo {author} {\bibfnamefont {S.}~\bibnamefont {Florens}},\ }\bibfield  {title} {\bibinfo {title} {{Particle production in ultrastrong-coupling waveguide QED}},\ }\href {https://doi.org/10.1103/physreva.98.043816} {\bibfield  {journal} {\bibinfo  {journal} {Physical Review A}\ }\textbf {\bibinfo {volume} {98}},\ \bibinfo {pages} {043816} (\bibinfo {year} {2018})}\BibitemShut {NoStop}%
\bibitem [{\citenamefont {Houzet}\ and\ \citenamefont {Glazman}(2020)}]{Houzet.2020}%
  \BibitemOpen
  \bibfield  {author} {\bibinfo {author} {\bibfnamefont {M.}~\bibnamefont {Houzet}}\ and\ \bibinfo {author} {\bibfnamefont {L.~I.}\ \bibnamefont {Glazman}},\ }\bibfield  {title} {\bibinfo {title} {{Critical Fluorescence of a Transmon at the Schmid Transition}},\ }\href {https://doi.org/10.1103/physrevlett.125.267701} {\bibfield  {journal} {\bibinfo  {journal} {Physical Review Letters}\ }\textbf {\bibinfo {volume} {125}},\ \bibinfo {pages} {267701} (\bibinfo {year} {2020})}\BibitemShut {NoStop}%
\bibitem [{\citenamefont {Burshtein}\ \emph {et~al.}(2021)\citenamefont {Burshtein}, \citenamefont {Kuzmin}, \citenamefont {Manucharyan},\ and\ \citenamefont {Goldstein}}]{Burshtein.2021}%
  \BibitemOpen
  \bibfield  {author} {\bibinfo {author} {\bibfnamefont {A.}~\bibnamefont {Burshtein}}, \bibinfo {author} {\bibfnamefont {R.}~\bibnamefont {Kuzmin}}, \bibinfo {author} {\bibfnamefont {V.~E.}\ \bibnamefont {Manucharyan}},\ and\ \bibinfo {author} {\bibfnamefont {M.}~\bibnamefont {Goldstein}},\ }\bibfield  {title} {\bibinfo {title} {{Photon-Instanton Collider Implemented by a Superconducting Circuit}},\ }\href {https://doi.org/10.1103/physrevlett.126.137701} {\bibfield  {journal} {\bibinfo  {journal} {Physical Review Letters}\ }\textbf {\bibinfo {volume} {126}},\ \bibinfo {pages} {137701} (\bibinfo {year} {2021})}\BibitemShut {NoStop}%
\bibitem [{\citenamefont {Burshtein}\ and\ \citenamefont {Goldstein}(2023)}]{burshtein2023inelastic}%
  \BibitemOpen
  \bibfield  {author} {\bibinfo {author} {\bibfnamefont {A.}~\bibnamefont {Burshtein}}\ and\ \bibinfo {author} {\bibfnamefont {M.}~\bibnamefont {Goldstein}},\ }\href@noop {} {\bibinfo {title} {Inelastic decay from integrability}} (\bibinfo {year} {2023}),\ \Eprint {https://arxiv.org/abs/2308.15542} {arXiv:2308.15542 [quant-ph]} \BibitemShut {NoStop}%
\bibitem [{\citenamefont {Houzet}\ \emph {et~al.}(2023)\citenamefont {Houzet}, \citenamefont {Yamamoto},\ and\ \citenamefont {Glazman}}]{houzet2023microwave}%
  \BibitemOpen
  \bibfield  {author} {\bibinfo {author} {\bibfnamefont {M.}~\bibnamefont {Houzet}}, \bibinfo {author} {\bibfnamefont {T.}~\bibnamefont {Yamamoto}},\ and\ \bibinfo {author} {\bibfnamefont {L.~I.}\ \bibnamefont {Glazman}},\ }\href@noop {} {\bibinfo {title} {Microwave spectroscopy of schmid transition}} (\bibinfo {year} {2023}),\ \Eprint {https://arxiv.org/abs/2308.16072} {arXiv:2308.16072 [cond-mat.supr-con]} \BibitemShut {NoStop}%
\bibitem [{\citenamefont {Mehta}\ \emph {et~al.}(2022)\citenamefont {Mehta}, \citenamefont {Ciuti}, \citenamefont {Kuzmin},\ and\ \citenamefont {Manucharyan}}]{Mehta.2022tcj}%
  \BibitemOpen
  \bibfield  {author} {\bibinfo {author} {\bibfnamefont {N.}~\bibnamefont {Mehta}}, \bibinfo {author} {\bibfnamefont {C.}~\bibnamefont {Ciuti}}, \bibinfo {author} {\bibfnamefont {R.}~\bibnamefont {Kuzmin}},\ and\ \bibinfo {author} {\bibfnamefont {V.~E.}\ \bibnamefont {Manucharyan}},\ }\bibfield  {title} {\bibinfo {title} {{Theory of strong down-conversion in multi-mode cavity and circuit QED}},\ }\bibfield  {journal} {\bibinfo  {journal} {arXiv}\ }\href {https://doi.org/10.48550/arxiv.2210.14681} {10.48550/arxiv.2210.14681} (\bibinfo {year} {2022})\BibitemShut {NoStop}%
\bibitem [{\citenamefont {Caldeira}\ and\ \citenamefont {Leggett}(1983)}]{Caldeira.1983}%
  \BibitemOpen
  \bibfield  {author} {\bibinfo {author} {\bibfnamefont {A.~O.}\ \bibnamefont {Caldeira}}\ and\ \bibinfo {author} {\bibfnamefont {A.~J.}\ \bibnamefont {Leggett}},\ }\bibfield  {title} {\bibinfo {title} {{Quantum tunnelling in a dissipative system}},\ }\href {https://doi.org/10.1016/0003-4916(83)90202-6} {\bibfield  {journal} {\bibinfo  {journal} {Annals of Physics}\ }\textbf {\bibinfo {volume} {149}},\ \bibinfo {pages} {374 } (\bibinfo {year} {1983})}\BibitemShut {NoStop}%
\bibitem [{\citenamefont {Bencheikh}\ \emph {et~al.}(2007)\citenamefont {Bencheikh}, \citenamefont {Gravier}, \citenamefont {Douady}, \citenamefont {Levenson},\ and\ \citenamefont {Boulanger}}]{Bencheikh.2007}%
  \BibitemOpen
  \bibfield  {author} {\bibinfo {author} {\bibfnamefont {K.}~\bibnamefont {Bencheikh}}, \bibinfo {author} {\bibfnamefont {F.}~\bibnamefont {Gravier}}, \bibinfo {author} {\bibfnamefont {J.}~\bibnamefont {Douady}}, \bibinfo {author} {\bibfnamefont {A.}~\bibnamefont {Levenson}},\ and\ \bibinfo {author} {\bibfnamefont {B.}~\bibnamefont {Boulanger}},\ }\bibfield  {title} {\bibinfo {title} {{Triple photons: a challenge in nonlinear and quantum optics}},\ }\href {https://doi.org/10.1016/j.crhy.2006.07.014} {\bibfield  {journal} {\bibinfo  {journal} {Comptes Rendus Physique}\ }\textbf {\bibinfo {volume} {8}},\ \bibinfo {pages} {206} (\bibinfo {year} {2007})}\BibitemShut {NoStop}%
\bibitem [{\citenamefont {Vrajitoarea}\ \emph {et~al.}(2022)\citenamefont {Vrajitoarea}, \citenamefont {Belyansky}, \citenamefont {Lundgren}, \citenamefont {Whitsitt}, \citenamefont {Gorshkov},\ and\ \citenamefont {Houck}}]{Vrajitoarea.2022}%
  \BibitemOpen
  \bibfield  {author} {\bibinfo {author} {\bibfnamefont {A.}~\bibnamefont {Vrajitoarea}}, \bibinfo {author} {\bibfnamefont {R.}~\bibnamefont {Belyansky}}, \bibinfo {author} {\bibfnamefont {R.}~\bibnamefont {Lundgren}}, \bibinfo {author} {\bibfnamefont {S.}~\bibnamefont {Whitsitt}}, \bibinfo {author} {\bibfnamefont {A.~V.}\ \bibnamefont {Gorshkov}},\ and\ \bibinfo {author} {\bibfnamefont {A.~A.}\ \bibnamefont {Houck}},\ }\bibfield  {title} {\bibinfo {title} {{Ultrastrong light-matter interaction in a photonic crystal}},\ }\href@noop {} {\bibfield  {journal} {\bibinfo  {journal} {arXiv}\ } (\bibinfo {year} {2022})},\ \Eprint {https://arxiv.org/abs/2209.14972} {2209.14972} \BibitemShut {NoStop}%
\bibitem [{\citenamefont {Kockum}\ \emph {et~al.}(2017)\citenamefont {Kockum}, \citenamefont {Macri}, \citenamefont {Garziano}, \citenamefont {Savasta},\ and\ \citenamefont {Nori}}]{Kockum2017}%
  \BibitemOpen
  \bibfield  {author} {\bibinfo {author} {\bibfnamefont {A.~F.}\ \bibnamefont {Kockum}}, \bibinfo {author} {\bibfnamefont {V.}~\bibnamefont {Macri}}, \bibinfo {author} {\bibfnamefont {L.}~\bibnamefont {Garziano}}, \bibinfo {author} {\bibfnamefont {S.}~\bibnamefont {Savasta}},\ and\ \bibinfo {author} {\bibfnamefont {F.}~\bibnamefont {Nori}},\ }\bibfield  {title} {\bibinfo {title} {{Frequency conversion in ultrastrong cavity QED}},\ }\href {https://doi.org/10.1038/s41598-017-04225-3} {\bibfield  {journal} {\bibinfo  {journal} {Scientific Reports}\ }\textbf {\bibinfo {volume} {7}},\ \bibinfo {pages} {5313} (\bibinfo {year} {2017})}\BibitemShut {NoStop}%
\bibitem [{\citenamefont {Koshino}\ \emph {et~al.}(2022)\citenamefont {Koshino}, \citenamefont {Shitara}, \citenamefont {Ao},\ and\ \citenamefont {Semba}}]{Koshino.2022}%
  \BibitemOpen
  \bibfield  {author} {\bibinfo {author} {\bibfnamefont {K.}~\bibnamefont {Koshino}}, \bibinfo {author} {\bibfnamefont {T.}~\bibnamefont {Shitara}}, \bibinfo {author} {\bibfnamefont {Z.}~\bibnamefont {Ao}},\ and\ \bibinfo {author} {\bibfnamefont {K.}~\bibnamefont {Semba}},\ }\bibfield  {title} {\bibinfo {title} {{Deterministic three-photon down-conversion by a passive ultrastrong cavity-QED system}},\ }\href {https://doi.org/10.1103/physrevresearch.4.013013} {\bibfield  {journal} {\bibinfo  {journal} {Physical Review Research}\ }\textbf {\bibinfo {volume} {4}},\ \bibinfo {pages} {013013} (\bibinfo {year} {2022})}\BibitemShut {NoStop}%
\bibitem [{\citenamefont {Magazzu}\ \emph {et~al.}(2018)\citenamefont {Magazzu}, \citenamefont {Forn-D{\'\i}az}, \citenamefont {Belyansky}, \citenamefont {Orgiazzi}, \citenamefont {Yurtalan}, \citenamefont {Otto}, \citenamefont {Lupa{\c s}cu}, \citenamefont {Wilson},\ and\ \citenamefont {Grifoni}}]{Magazzu.2018}%
  \BibitemOpen
  \bibfield  {author} {\bibinfo {author} {\bibfnamefont {L.}~\bibnamefont {Magazzu}}, \bibinfo {author} {\bibfnamefont {P.}~\bibnamefont {Forn-D{\'\i}az}}, \bibinfo {author} {\bibfnamefont {R.}~\bibnamefont {Belyansky}}, \bibinfo {author} {\bibfnamefont {J.~L.}\ \bibnamefont {Orgiazzi}}, \bibinfo {author} {\bibfnamefont {M.~A.}\ \bibnamefont {Yurtalan}}, \bibinfo {author} {\bibfnamefont {M.~R.}\ \bibnamefont {Otto}}, \bibinfo {author} {\bibfnamefont {A.}~\bibnamefont {Lupa{\c s}cu}}, \bibinfo {author} {\bibfnamefont {C.~M.}\ \bibnamefont {Wilson}},\ and\ \bibinfo {author} {\bibfnamefont {M.}~\bibnamefont {Grifoni}},\ }\bibfield  {title} {\bibinfo {title} {{Probing the strongly driven spin-boson model in a superconducting quantum circuit}},\ }\href {https://doi.org/10.1038/s41467-018-03626-w} {\bibfield  {journal} {\bibinfo  {journal} {Nature Communications}\ }\textbf {\bibinfo {volume} {9}},\ \bibinfo {pages} {1403} (\bibinfo {year} {2018})}\BibitemShut {NoStop}%
\bibitem [{\citenamefont {Zakka-Bajjani}\ \emph {et~al.}(2011)\citenamefont {Zakka-Bajjani}, \citenamefont {Nguyen}, \citenamefont {Lee}, \citenamefont {Vale}, \citenamefont {Simmonds},\ and\ \citenamefont {Aumentado}}]{Zakka-Bajjani.2011pod}%
  \BibitemOpen
  \bibfield  {author} {\bibinfo {author} {\bibfnamefont {E.}~\bibnamefont {Zakka-Bajjani}}, \bibinfo {author} {\bibfnamefont {F.}~\bibnamefont {Nguyen}}, \bibinfo {author} {\bibfnamefont {M.}~\bibnamefont {Lee}}, \bibinfo {author} {\bibfnamefont {L.~R.}\ \bibnamefont {Vale}}, \bibinfo {author} {\bibfnamefont {R.~W.}\ \bibnamefont {Simmonds}},\ and\ \bibinfo {author} {\bibfnamefont {J.}~\bibnamefont {Aumentado}},\ }\bibfield  {title} {\bibinfo {title} {{Quantum superposition of a single microwave photon in two different 'colour' states}},\ }\href {https://doi.org/10.1038/nphys2035} {\bibfield  {journal} {\bibinfo  {journal} {Nature Physics}\ }\textbf {\bibinfo {volume} {7}},\ \bibinfo {pages} {599} (\bibinfo {year} {2011})}\BibitemShut {NoStop}%
\bibitem [{\citenamefont {Svensson}\ \emph {et~al.}(2018)\citenamefont {Svensson}, \citenamefont {Bengtsson}, \citenamefont {Bylander}, \citenamefont {Shumeiko},\ and\ \citenamefont {Delsing}}]{Svensson.2018kch}%
  \BibitemOpen
  \bibfield  {author} {\bibinfo {author} {\bibfnamefont {I.-M.}\ \bibnamefont {Svensson}}, \bibinfo {author} {\bibfnamefont {A.}~\bibnamefont {Bengtsson}}, \bibinfo {author} {\bibfnamefont {J.}~\bibnamefont {Bylander}}, \bibinfo {author} {\bibfnamefont {V.}~\bibnamefont {Shumeiko}},\ and\ \bibinfo {author} {\bibfnamefont {P.}~\bibnamefont {Delsing}},\ }\bibfield  {title} {\bibinfo {title} {{Period multiplication in a parametrically driven superconducting resonator}},\ }\href {https://doi.org/10.1063/1.5026974} {\bibfield  {journal} {\bibinfo  {journal} {Applied Physics Letters}\ }\textbf {\bibinfo {volume} {113}},\ \bibinfo {pages} {022602} (\bibinfo {year} {2018})}\BibitemShut {NoStop}%
\bibitem [{\citenamefont {Chang}\ \emph {et~al.}(2020)\citenamefont {Chang}, \citenamefont {Sabin}, \citenamefont {Forn-D{\'\i}az}, \citenamefont {Quijandr{\'\i}a}, \citenamefont {Vadiraj}, \citenamefont {Nsanzineza}, \citenamefont {Johansson},\ and\ \citenamefont {Wilson}}]{Chang.2020740z}%
  \BibitemOpen
  \bibfield  {author} {\bibinfo {author} {\bibfnamefont {C.~W.~S.}\ \bibnamefont {Chang}}, \bibinfo {author} {\bibfnamefont {C.}~\bibnamefont {Sabin}}, \bibinfo {author} {\bibfnamefont {P.}~\bibnamefont {Forn-D{\'\i}az}}, \bibinfo {author} {\bibfnamefont {F.}~\bibnamefont {Quijandr{\'\i}a}}, \bibinfo {author} {\bibfnamefont {A.~M.}\ \bibnamefont {Vadiraj}}, \bibinfo {author} {\bibfnamefont {I.}~\bibnamefont {Nsanzineza}}, \bibinfo {author} {\bibfnamefont {G.}~\bibnamefont {Johansson}},\ and\ \bibinfo {author} {\bibfnamefont {C.~M.}\ \bibnamefont {Wilson}},\ }\bibfield  {title} {\bibinfo {title} {{Observation of Three-Photon Spontaneous Parametric Down-Conversion in a Superconducting Parametric Cavity}},\ }\href {https://doi.org/10.1103/physrevx.10.011011} {\bibfield  {journal} {\bibinfo  {journal} {Physical Review X}\ }\textbf {\bibinfo {volume} {10}},\ \bibinfo {pages} {011011} (\bibinfo {year} {2020})}\BibitemShut {NoStop}%
\bibitem [{\citenamefont {Gasparinetti}\ \emph {et~al.}(2017)\citenamefont {Gasparinetti}, \citenamefont {Pechal}, \citenamefont {Besse}, \citenamefont {Mondal}, \citenamefont {Eichler},\ and\ \citenamefont {Wallraff}}]{Gasparinetti.2017}%
  \BibitemOpen
  \bibfield  {author} {\bibinfo {author} {\bibfnamefont {S.}~\bibnamefont {Gasparinetti}}, \bibinfo {author} {\bibfnamefont {M.}~\bibnamefont {Pechal}}, \bibinfo {author} {\bibfnamefont {J.-C.}\ \bibnamefont {Besse}}, \bibinfo {author} {\bibfnamefont {M.}~\bibnamefont {Mondal}}, \bibinfo {author} {\bibfnamefont {C.}~\bibnamefont {Eichler}},\ and\ \bibinfo {author} {\bibfnamefont {A.}~\bibnamefont {Wallraff}},\ }\bibfield  {title} {\bibinfo {title} {{Correlations and Entanglement of Microwave Photons Emitted in a Cascade Decay}},\ }\href {https://doi.org/10.1103/physrevlett.119.140504} {\bibfield  {journal} {\bibinfo  {journal} {Physical Review Letters}\ }\textbf {\bibinfo {volume} {119}},\ \bibinfo {pages} {140504 } (\bibinfo {year} {2017})}\BibitemShut {NoStop}%
\bibitem [{\citenamefont {Rieger}\ \emph {et~al.}(2023)\citenamefont {Rieger}, \citenamefont {G\"unzler}, \citenamefont {Spiecker}, \citenamefont {Nambisan}, \citenamefont {Wernsdorfer},\ and\ \citenamefont {Pop}}]{Rieger.20225gr}%
  \BibitemOpen
  \bibfield  {author} {\bibinfo {author} {\bibfnamefont {D.}~\bibnamefont {Rieger}}, \bibinfo {author} {\bibfnamefont {S.}~\bibnamefont {G\"unzler}}, \bibinfo {author} {\bibfnamefont {M.}~\bibnamefont {Spiecker}}, \bibinfo {author} {\bibfnamefont {A.}~\bibnamefont {Nambisan}}, \bibinfo {author} {\bibfnamefont {W.}~\bibnamefont {Wernsdorfer}},\ and\ \bibinfo {author} {\bibfnamefont {I.}~\bibnamefont {Pop}},\ }\bibfield  {title} {\bibinfo {title} {Fano interference in microwave resonator measurements},\ }\href {https://doi.org/10.1103/PhysRevApplied.20.014059} {\bibfield  {journal} {\bibinfo  {journal} {Phys. Rev. Appl.}\ }\textbf {\bibinfo {volume} {20}},\ \bibinfo {pages} {014059} (\bibinfo {year} {2023})}\BibitemShut {NoStop}%
\bibitem [{\citenamefont {Ranadive}\ \emph {et~al.}(2022)\citenamefont {Ranadive}, \citenamefont {Esposito}, \citenamefont {Planat}, \citenamefont {Bonet}, \citenamefont {Naud}, \citenamefont {Buisson}, \citenamefont {Guichard},\ and\ \citenamefont {Roch}}]{Ranadive.2022}%
  \BibitemOpen
  \bibfield  {author} {\bibinfo {author} {\bibfnamefont {A.}~\bibnamefont {Ranadive}}, \bibinfo {author} {\bibfnamefont {M.}~\bibnamefont {Esposito}}, \bibinfo {author} {\bibfnamefont {L.}~\bibnamefont {Planat}}, \bibinfo {author} {\bibfnamefont {E.}~\bibnamefont {Bonet}}, \bibinfo {author} {\bibfnamefont {C.}~\bibnamefont {Naud}}, \bibinfo {author} {\bibfnamefont {O.}~\bibnamefont {Buisson}}, \bibinfo {author} {\bibfnamefont {W.}~\bibnamefont {Guichard}},\ and\ \bibinfo {author} {\bibfnamefont {N.}~\bibnamefont {Roch}},\ }\bibfield  {title} {\bibinfo {title} {{Kerr reversal in Josephson meta-material and traveling wave parametric amplification}},\ }\href {https://doi.org/10.1038/s41467-022-29375-5} {\bibfield  {journal} {\bibinfo  {journal} {Nature Communications}\ }\textbf {\bibinfo {volume} {13}},\ \bibinfo {pages} {1737} (\bibinfo {year} {2022})}\BibitemShut {NoStop}%
\bibitem [{\citenamefont {Mollow}(1969)}]{Mollow.1969}%
  \BibitemOpen
  \bibfield  {author} {\bibinfo {author} {\bibfnamefont {B.~R.}\ \bibnamefont {Mollow}},\ }\bibfield  {title} {\bibinfo {title} {Power spectrum of light scattered by two-level systems},\ }\href {https://doi.org/10.1103/PhysRev.188.1969} {\bibfield  {journal} {\bibinfo  {journal} {Phys. Rev.}\ }\textbf {\bibinfo {volume} {188}},\ \bibinfo {pages} {1969} (\bibinfo {year} {1969})}\BibitemShut {NoStop}%
\bibitem [{\citenamefont {Hriscu}\ and\ \citenamefont {Nazarov}(2011)}]{Hriscu.2011}%
  \BibitemOpen
  \bibfield  {author} {\bibinfo {author} {\bibfnamefont {A.~M.}\ \bibnamefont {Hriscu}}\ and\ \bibinfo {author} {\bibfnamefont {Y.~V.}\ \bibnamefont {Nazarov}},\ }\bibfield  {title} {\bibinfo {title} {Model of a proposed superconducting phase slip oscillator: A method for obtaining few-photon nonlinearities},\ }\href {https://doi.org/10.1103/PhysRevLett.106.077004} {\bibfield  {journal} {\bibinfo  {journal} {Phys. Rev. Lett.}\ }\textbf {\bibinfo {volume} {106}},\ \bibinfo {pages} {077004} (\bibinfo {year} {2011})}\BibitemShut {NoStop}%
\bibitem [{\citenamefont {Smith}\ \emph {et~al.}(2023)\citenamefont {Smith}, \citenamefont {Borgognoni}, \citenamefont {Villiers}, \citenamefont {Roverc'h}, \citenamefont {Palomo}, \citenamefont {Delbecq}, \citenamefont {Kontos}, \citenamefont {Campagne-Ibarcq}, \citenamefont {Douçot},\ and\ \citenamefont {Leghtas}}]{Smith.2023}%
  \BibitemOpen
  \bibfield  {author} {\bibinfo {author} {\bibfnamefont {W.~C.}\ \bibnamefont {Smith}}, \bibinfo {author} {\bibfnamefont {A.}~\bibnamefont {Borgognoni}}, \bibinfo {author} {\bibfnamefont {M.}~\bibnamefont {Villiers}}, \bibinfo {author} {\bibfnamefont {E.}~\bibnamefont {Roverc'h}}, \bibinfo {author} {\bibfnamefont {J.}~\bibnamefont {Palomo}}, \bibinfo {author} {\bibfnamefont {M.~R.}\ \bibnamefont {Delbecq}}, \bibinfo {author} {\bibfnamefont {T.}~\bibnamefont {Kontos}}, \bibinfo {author} {\bibfnamefont {P.}~\bibnamefont {Campagne-Ibarcq}}, \bibinfo {author} {\bibfnamefont {B.}~\bibnamefont {Douçot}},\ and\ \bibinfo {author} {\bibfnamefont {Z.}~\bibnamefont {Leghtas}},\ }\bibfield  {title} {\bibinfo {title} {{Spectral signature of high-order photon processes mediated by Cooper-pair pairing}},\ }\bibfield  {journal} {\bibinfo  {journal} {arXiv}\ }\href {https://doi.org/10.48550/arxiv.2312.15075} {10.48550/arxiv.2312.15075} (\bibinfo {year} {2023}),\ \Eprint {https://arxiv.org/abs/2312.15075} {2312.15075}
  \BibitemShut {NoStop}%
\bibitem [{\citenamefont {Somoroff}\ \emph {et~al.}(2023)\citenamefont {Somoroff}, \citenamefont {Ficheux}, \citenamefont {Mencia}, \citenamefont {Xiong}, \citenamefont {Kuzmin},\ and\ \citenamefont {Manucharyan}}]{Somoroff.2021}%
  \BibitemOpen
  \bibfield  {author} {\bibinfo {author} {\bibfnamefont {A.}~\bibnamefont {Somoroff}}, \bibinfo {author} {\bibfnamefont {Q.}~\bibnamefont {Ficheux}}, \bibinfo {author} {\bibfnamefont {R.~A.}\ \bibnamefont {Mencia}}, \bibinfo {author} {\bibfnamefont {H.}~\bibnamefont {Xiong}}, \bibinfo {author} {\bibfnamefont {R.}~\bibnamefont {Kuzmin}},\ and\ \bibinfo {author} {\bibfnamefont {V.~E.}\ \bibnamefont {Manucharyan}},\ }\bibfield  {title} {\bibinfo {title} {Millisecond coherence in a superconducting qubit},\ }\href {https://doi.org/10.1103/PhysRevLett.130.267001} {\bibfield  {journal} {\bibinfo  {journal} {Phys. Rev. Lett.}\ }\textbf {\bibinfo {volume} {130}},\ \bibinfo {pages} {267001} (\bibinfo {year} {2023})}\BibitemShut {NoStop}%
\bibitem [{\citenamefont {Gyenis}\ \emph {et~al.}(2021)\citenamefont {Gyenis}, \citenamefont {Mundada}, \citenamefont {Di~Paolo}, \citenamefont {Hazard}, \citenamefont {You}, \citenamefont {Schuster}, \citenamefont {Koch}, \citenamefont {Blais},\ and\ \citenamefont {Houck}}]{Gyenis.2021}%
  \BibitemOpen
  \bibfield  {author} {\bibinfo {author} {\bibfnamefont {A.}~\bibnamefont {Gyenis}}, \bibinfo {author} {\bibfnamefont {P.~S.}\ \bibnamefont {Mundada}}, \bibinfo {author} {\bibfnamefont {A.}~\bibnamefont {Di~Paolo}}, \bibinfo {author} {\bibfnamefont {T.~M.}\ \bibnamefont {Hazard}}, \bibinfo {author} {\bibfnamefont {X.}~\bibnamefont {You}}, \bibinfo {author} {\bibfnamefont {D.~I.}\ \bibnamefont {Schuster}}, \bibinfo {author} {\bibfnamefont {J.}~\bibnamefont {Koch}}, \bibinfo {author} {\bibfnamefont {A.}~\bibnamefont {Blais}},\ and\ \bibinfo {author} {\bibfnamefont {A.~A.}\ \bibnamefont {Houck}},\ }\bibfield  {title} {\bibinfo {title} {Experimental realization of a protected superconducting circuit derived from the $0$--$\ensuremath{\pi}$ qubit},\ }\href {https://doi.org/10.1103/PRXQuantum.2.010339} {\bibfield  {journal} {\bibinfo  {journal} {PRX Quantum}\ }\textbf {\bibinfo {volume} {2}},\ \bibinfo {pages} {010339} (\bibinfo {year} {2021})}\BibitemShut {NoStop}%
\bibitem [{\citenamefont {Mizel}\ and\ \citenamefont {Yanay}(2020)}]{Mizel.2019}%
  \BibitemOpen
  \bibfield  {author} {\bibinfo {author} {\bibfnamefont {A.}~\bibnamefont {Mizel}}\ and\ \bibinfo {author} {\bibfnamefont {Y.}~\bibnamefont {Yanay}},\ }\bibfield  {title} {\bibinfo {title} {Right-sizing fluxonium against charge noise},\ }\href {https://doi.org/10.1103/PhysRevB.102.014512} {\bibfield  {journal} {\bibinfo  {journal} {Phys. Rev. B}\ }\textbf {\bibinfo {volume} {102}},\ \bibinfo {pages} {014512} (\bibinfo {year} {2020})}\BibitemShut {NoStop}%
\bibitem [{\citenamefont {Di~Paolo}\ \emph {et~al.}(2021)\citenamefont {Di~Paolo}, \citenamefont {Baker}, \citenamefont {Foley}, \citenamefont {S{\'e}n{\'e}chal},\ and\ \citenamefont {Blais}}]{Paolo.2019}%
  \BibitemOpen
  \bibfield  {author} {\bibinfo {author} {\bibfnamefont {A.}~\bibnamefont {Di~Paolo}}, \bibinfo {author} {\bibfnamefont {T.~E.}\ \bibnamefont {Baker}}, \bibinfo {author} {\bibfnamefont {A.}~\bibnamefont {Foley}}, \bibinfo {author} {\bibfnamefont {D.}~\bibnamefont {S{\'e}n{\'e}chal}},\ and\ \bibinfo {author} {\bibfnamefont {A.}~\bibnamefont {Blais}},\ }\bibfield  {title} {\bibinfo {title} {Efficient modeling of superconducting quantum circuits with tensor networks},\ }\href {https://doi.org/10.1038/s41534-020-00352-4} {\bibfield  {journal} {\bibinfo  {journal} {npj Quantum Information}\ }\textbf {\bibinfo {volume} {7}},\ \bibinfo {pages} {11} (\bibinfo {year} {2021})}\BibitemShut {NoStop}%
\bibitem [{\citenamefont {Kaur}\ \emph {et~al.}(2021)\citenamefont {Kaur}, \citenamefont {S{\'e}pulcre}, \citenamefont {Roch}, \citenamefont {Snyman}, \citenamefont {Florens},\ and\ \citenamefont {Bera}}]{10.1103/physrevlett.127.237702}%
  \BibitemOpen
  \bibfield  {author} {\bibinfo {author} {\bibfnamefont {K.}~\bibnamefont {Kaur}}, \bibinfo {author} {\bibfnamefont {T.}~\bibnamefont {S{\'e}pulcre}}, \bibinfo {author} {\bibfnamefont {N.}~\bibnamefont {Roch}}, \bibinfo {author} {\bibfnamefont {I.}~\bibnamefont {Snyman}}, \bibinfo {author} {\bibfnamefont {S.}~\bibnamefont {Florens}},\ and\ \bibinfo {author} {\bibfnamefont {S.}~\bibnamefont {Bera}},\ }\bibfield  {title} {\bibinfo {title} {{Spin-Boson Quantum Phase Transition in Multilevel Superconducting Qubits}},\ }\href {https://doi.org/10.1103/physrevlett.127.237702} {\bibfield  {journal} {\bibinfo  {journal} {Physical Review Letters}\ }\textbf {\bibinfo {volume} {127}},\ \bibinfo {pages} {237702} (\bibinfo {year} {2021})}\BibitemShut {NoStop}%
\bibitem [{\citenamefont {Marquet}\ \emph {et~al.}(2023)\citenamefont {Marquet}, \citenamefont {Essig}, \citenamefont {Cohen}, \citenamefont {Cottet}, \citenamefont {Murani}, \citenamefont {Abertinale}, \citenamefont {Dupouy}, \citenamefont {Bienfait}, \citenamefont {Peronnin}, \citenamefont {Jezouin}, \citenamefont {Lescanne},\ and\ \citenamefont {Huard}}]{Marquet.2023}%
  \BibitemOpen
  \bibfield  {author} {\bibinfo {author} {\bibfnamefont {A.}~\bibnamefont {Marquet}}, \bibinfo {author} {\bibfnamefont {A.}~\bibnamefont {Essig}}, \bibinfo {author} {\bibfnamefont {J.}~\bibnamefont {Cohen}}, \bibinfo {author} {\bibfnamefont {N.}~\bibnamefont {Cottet}}, \bibinfo {author} {\bibfnamefont {A.}~\bibnamefont {Murani}}, \bibinfo {author} {\bibfnamefont {E.}~\bibnamefont {Abertinale}}, \bibinfo {author} {\bibfnamefont {S.}~\bibnamefont {Dupouy}}, \bibinfo {author} {\bibfnamefont {A.}~\bibnamefont {Bienfait}}, \bibinfo {author} {\bibfnamefont {T.}~\bibnamefont {Peronnin}}, \bibinfo {author} {\bibfnamefont {S.}~\bibnamefont {Jezouin}}, \bibinfo {author} {\bibfnamefont {R.}~\bibnamefont {Lescanne}},\ and\ \bibinfo {author} {\bibfnamefont {B.}~\bibnamefont {Huard}},\ }\bibfield  {title} {\bibinfo {title} {{Autoparametric resonance extending the bit-flip time of a cat qubit up to 0.3 s}},\ }\bibfield  {journal} {\bibinfo  {journal} {arXiv}\ }\href {https://doi.org/10.48550/arxiv.2307.06761}
  {10.48550/arxiv.2307.06761} (\bibinfo {year} {2023})\BibitemShut {NoStop}%
\bibitem [{\citenamefont {Silva}\ \emph {et~al.}(2010)\citenamefont {Silva}, \citenamefont {Bozyigit}, \citenamefont {Wallraff},\ and\ \citenamefont {Blais}}]{Silva.2010}%
  \BibitemOpen
  \bibfield  {author} {\bibinfo {author} {\bibfnamefont {M.~D.}\ \bibnamefont {Silva}}, \bibinfo {author} {\bibfnamefont {D.}~\bibnamefont {Bozyigit}}, \bibinfo {author} {\bibfnamefont {A.}~\bibnamefont {Wallraff}},\ and\ \bibinfo {author} {\bibfnamefont {A.}~\bibnamefont {Blais}},\ }\bibfield  {title} {\bibinfo {title} {{Schemes for the observation of photon correlation functions in circuit QED with linear detectors}},\ }\href {https://doi.org/10.1103/physreva.82.043804} {\bibfield  {journal} {\bibinfo  {journal} {Physical Review A}\ }\textbf {\bibinfo {volume} {82}},\ \bibinfo {pages} {043804} (\bibinfo {year} {2010})}\BibitemShut {NoStop}%
\bibitem [{\citenamefont {Peropadre}\ \emph {et~al.}(2013)\citenamefont {Peropadre}, \citenamefont {Zueco}, \citenamefont {Porras},\ and\ \citenamefont {Garcia-Ripoll}}]{Peropadre.2013}%
  \BibitemOpen
  \bibfield  {author} {\bibinfo {author} {\bibfnamefont {B.}~\bibnamefont {Peropadre}}, \bibinfo {author} {\bibfnamefont {D.}~\bibnamefont {Zueco}}, \bibinfo {author} {\bibfnamefont {D.}~\bibnamefont {Porras}},\ and\ \bibinfo {author} {\bibfnamefont {J.}~\bibnamefont {Garcia-Ripoll}},\ }\bibfield  {title} {\bibinfo {title} {{Nonequilibrium and Nonperturbative Dynamics of Ultrastrong Coupling in Open Lines}},\ }\href {https://doi.org/10.1103/physrevlett.111.243602} {\bibfield  {journal} {\bibinfo  {journal} {Physical Review Letters}\ }\textbf {\bibinfo {volume} {111}},\ \bibinfo {pages} {243602} (\bibinfo {year} {2013})}\BibitemShut {NoStop}%
\bibitem [{\citenamefont {Wei\ss{}l}\ \emph {et~al.}(2015)\citenamefont {Wei\ss{}l}, \citenamefont {K\"ung}, \citenamefont {Dumur}, \citenamefont {Feofanov}, \citenamefont {Matei}, \citenamefont {Naud}, \citenamefont {Buisson}, \citenamefont {Hekking},\ and\ \citenamefont {Guichard}}]{weissl2015Kerr}%
  \BibitemOpen
  \bibfield  {author} {\bibinfo {author} {\bibfnamefont {T.}~\bibnamefont {Wei\ss{}l}}, \bibinfo {author} {\bibfnamefont {B.}~\bibnamefont {K\"ung}}, \bibinfo {author} {\bibfnamefont {E.}~\bibnamefont {Dumur}}, \bibinfo {author} {\bibfnamefont {A.~K.}\ \bibnamefont {Feofanov}}, \bibinfo {author} {\bibfnamefont {I.}~\bibnamefont {Matei}}, \bibinfo {author} {\bibfnamefont {C.}~\bibnamefont {Naud}}, \bibinfo {author} {\bibfnamefont {O.}~\bibnamefont {Buisson}}, \bibinfo {author} {\bibfnamefont {F.~W.~J.}\ \bibnamefont {Hekking}},\ and\ \bibinfo {author} {\bibfnamefont {W.}~\bibnamefont {Guichard}},\ }\bibfield  {title} {\bibinfo {title} {Kerr coefficients of plasma resonances in josephson junction chains},\ }\href {https://doi.org/10.1103/PhysRevB.92.104508} {\bibfield  {journal} {\bibinfo  {journal} {Phys. Rev. B}\ }\textbf {\bibinfo {volume} {92}},\ \bibinfo {pages} {104508} (\bibinfo {year} {2015})}\BibitemShut {NoStop}%
\end{thebibliography}%

%% RESET COUNTERS FOR APPENDED SUPPLEMENTARY MATERIAL
\setcounter{figure}{0}
\setcounter{table}{0}
\setcounter{equation}{0}

\global\long\def\theequation{S\arabic{equation}}
\global\long\def\thefigure{S\arabic{figure}}
\renewcommand{\thetable}{S\arabic{table}} 
\renewcommand{\arraystretch}{0.6}

\normalsize

\vspace{1.0cm}
\begin{center}
{\bf \large Supplementary information for
``Direct detection of down-converted photons spontaneously produced at a single Josephson junction''}
\end{center}

\subsection{Fit of the dispersion relation}
\label{app:modes}

The sample consists of a superconducting array made of 200 Josephson junctions (JJs) that are galvanically coupled to the measurement line, and terminated by a small (non-linear) junction. The junctions of the array are designed to be in the linear regime and the room temperature resistance measurements on test structures provide an estimated ratio of $E_\mathrm{J}^\mathrm{arr}/E_\mathrm{C}^\mathrm{arr} \simeq 284$ at zero flux. The array is actually implemented using asymmetric SQUIDs, (see Figure~\ref{fig1}(a), with the array junctions highlighted in light blue) with a large
asymmetry $d$ estimated at 0.75, which provides a slow modulation of the array characteristic impedance $Z_\mathrm{A}$ by means of an external magnetic flux $\Phi_\mathrm{B}$. The small JJ terminating the array  is implemented by a nearly symmetric SQUID (asymmetry parameter $d \simeq 0.025$) with bare Josephson energy $E_\mathrm{J}(\Phi_\mathrm{B}=0)/h = 27$ GHz. Its intrinsic capacitance is estimated as $C_\mathrm{J} \simeq 8.5$ fF leading to a ratio $E_\mathrm{J}/E_\mathrm{C} \simeq 3$. Figure~\ref{fig1}(a) shows a full equivalent circuit diagram of the device, as well and a scanning electron microscope (SEM) image of the measured sample. Table~\ref{Table1} below gives a summary the estimated circuit parameters of the device under investigation.

\begin{figure*}[htb]
\begin{center}
\includegraphics[width = 0.9\textwidth]{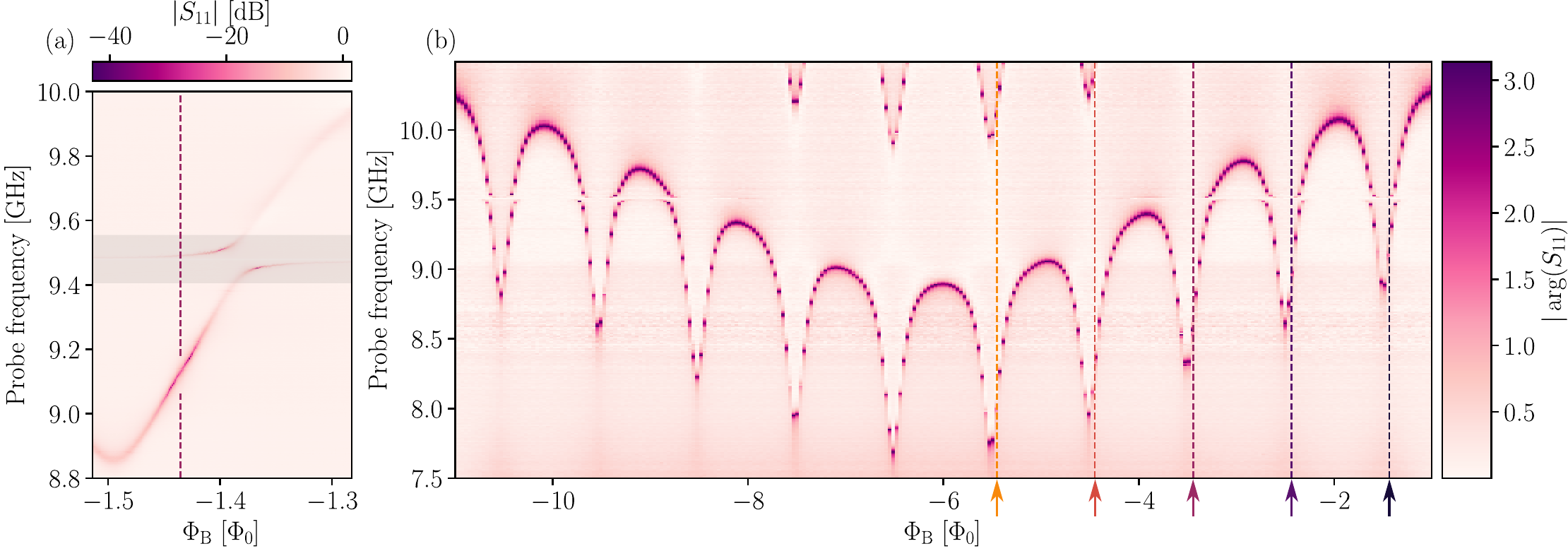}
\caption{\textbf{(a)} Color map of the measured magnitude of the reflection coefficient $\left |S_{11}(\omega_\text{d}) \right |$ of mode $\displaystyle \hat{b}$ versus drive frequency $\omega_\text{d}$ and external magnetic flux $\Phi_\mathrm{B}$ threading the small JJ loop. This close-up shows the arch where the photon emission was measured (see figure 3 of the main text). \textbf{(b)} Single-tone spectroscopy of mode $b$ on a wider range  (for the external flux $\Phi_\mathrm{B}$ varying between $-11\,\Phi_0$ and $-\Phi_0$), revealing two periods of modulation, corresponding to the  SQUIDs in the array (slow period) and in the small junction (fast period). The coupling of mode $b$ to a fixed parasitic mode around 9.5~GHz can be observed. 
%The horizontal line indicates the fixed value of mode $b$  frequency $\omega_b/2 \pi=9.13$ GHz% 
The five vertical lines indicate the flux values where a strong conversion peak is seen in Fig.~\ref{fig:FigAppB2}.}
\label{fig:FigAppB1}
\end{center}
\end{figure*}

\begin{table}[b]
\centering
\begin{tabular}{ccc}
\hline
Parameters & Boundary junction  & Chain junctions \\ 
\hline
$C_\mathrm{J}$ [fF] & 8.5 & 88 \\
$C_\mathrm{g}$ [fF] & & 0.7 \\
$E_\mathrm{J} /h$ [GHz] & 27.2 & 247.0 \\
$E_\mathrm{C}/h$ [GHz] & 9.1 & 0.9 \\
$E_\mathrm{J}/E_\mathrm{C}$ & 3.0 & 284.0 \\
$Z_\text{A}/R_\mathrm{Q}$ & 0.1 & 0.1 \\
$\omega_\mathrm{P}/2\pi$ [GHz] & 22.3 & 20.8 \\
\hline
\end{tabular}
\caption{\label{Table1} \small
 Characteristics of the junction array and the impurity for the device under
investigation.} 
\end{table}

Due to the impedance mismatch between the measurement line and the array, there is always a finite coupling between the measurement line and the system due to the non-zero reflection coefficient $(Z_0-
Z_\mathrm{A})/(Z_0+Z_\mathrm{A})$ at the left boundary. Using the ABCD matrix technique, it can be determined that, given the characteristic impedance of the array $Z_\mathrm{A}$, the galvanic coupling to the measurement line is sufficient to satisfy the condition $\Gamma_\mathrm{nr} \gtrsim g_{1\leftrightarrow3}/\hbar$. Given the boundary conditions, and approximating the 50 $\Omega$ measurement line by a short to ground (this is justified because the impedance of the measurement line, $Z_0 = 50\:\Omega$, is small compared to the characteristic impedance of the array, $Z_\mathrm{A} \simeq 910$ $\Omega$), wave vectors of the normal modes of the system are given by:
\begin{equation}
\label{wavevector}
\kappa_k=\frac{(k+1/2)\pi}{N-1/2} + \frac{\theta_k}{N-1/2},
\end{equation}
where $N$ is the number of junctions in the array. The first term on the right-hand side of Eq.~(\ref{wavevector}) is the unperturbed wave vector of a $\lambda/2$ lumped transmission line (with ground boundary conditions at both ends), while the second term originates from the impurity induced phase shift $\theta_k$, self-consistently determined by:
\begin{eqnarray}
\label{phaseshift}
    \theta_k &=& \arctan\left[
    \frac{\lambda_k-1}{\lambda_k+1}\sqrt{\left(\frac{4C}{C_\mathrm{g}}+1\right)\left(\frac{\omega_P^2}{\omega_k^2}+1\right)}\right],\\
    \lambda_k &=& 1 + \frac{L}{L_\textrm{J}}\frac{1-\omega_k^2 L_\textrm{J}\left(C_\textrm{J} + C_\textrm{g}\right)}{1-\omega_k^2LC},
\end{eqnarray}
with $\omega_\mathrm{P} = 1/\sqrt{LC}$ the array plasma frequency.

In order to characterise the sample under study, we measured the dispersion relation of the
system at zero flux $\Phi_\mathrm{B}=0$ using a technique known as two-tone spectroscopy~\cite{weissl2015Kerr}.
This technique makes it possible to measure the resonance frequency of modes outside
of the measurement bandwidth of the experimental setup by exploiting the cross-Kerr
effect. The idea is to probe the system in reflection at a single frequency corresponding
to the resonance frequency of a given mode and to drive the system with a second tone of
varying frequency. Each time the second tone drives a resonant mode of the system, the
resonant frequency of the probe mode gets shifted due to the Kerr interaction, resulting in a significant variation in the amplitude and phase of the reflection coefficient. It is the changes in the reflection of
the probe tone correlated with the frequency of the second tone that allow the dispersion
relation of the system to be extracted on the full frequency range. Figure~\ref{fig1}(b) shows the result of the two-tone
spectroscopy of the sample at flux $\Phi_\mathrm{B} = 0$. The analytical dispersion relation, indicated as light blue line in Fig.~\ref{fig1}(b) and given by 
\begin{equation}
    \omega_k =
    \sqrt{\frac{2(1-\cos ka)}{LC_\mathrm{g}+2LC(1-\cos ka)}},
\end{equation}
is fitted to the data by self-consistently solving for the wave vectors as given
by equations~(\ref{wavevector}) and (\ref{phaseshift}). The array ground capacitance $C_\mathrm{g} = 0.73$ fF is estimated
from this fitting procedure.

\subsection{Losses at different values of the chain impedance}
\label{app:impedance}

In the main text, we only mentioned the tunability of the small junction for a fixed value of the array impedance. The system can also be tuned using another knob, namely changing the array inductance by applying an external magnetic flux through the array SQUIDs. The external magnetic field used to tune the array SQUIDs is actually the same as that used to tune $E_\mathrm{J}$ of the small junction. The tuning of the two types of junctions is made essentially independent by choosing a small junction loop that is about fourteen times larger than the bigger loops of the array junctions (see Figure~\ref{fig1}), so that small variations of the magnetic field affect mainly the small junction, in the first approximation. 
Thus, for the same field, the flux threading the array SQUIDs is $\Phi_\mathrm{A} \simeq \Phi_\mathrm{B} /14$. 
Figure~\ref{fig:FigAppB1} shows the single-tone spectroscopy of mode $b$ for a large span of externally applied magnetic flux, revealing two distinct modulation periods associated with
the modulation of both the impurity junction (short periodicity) and the array junctions  (long periodicity).

\begin{figure}[htb]
\begin{center}
\includegraphics[width = 0.47\textwidth]{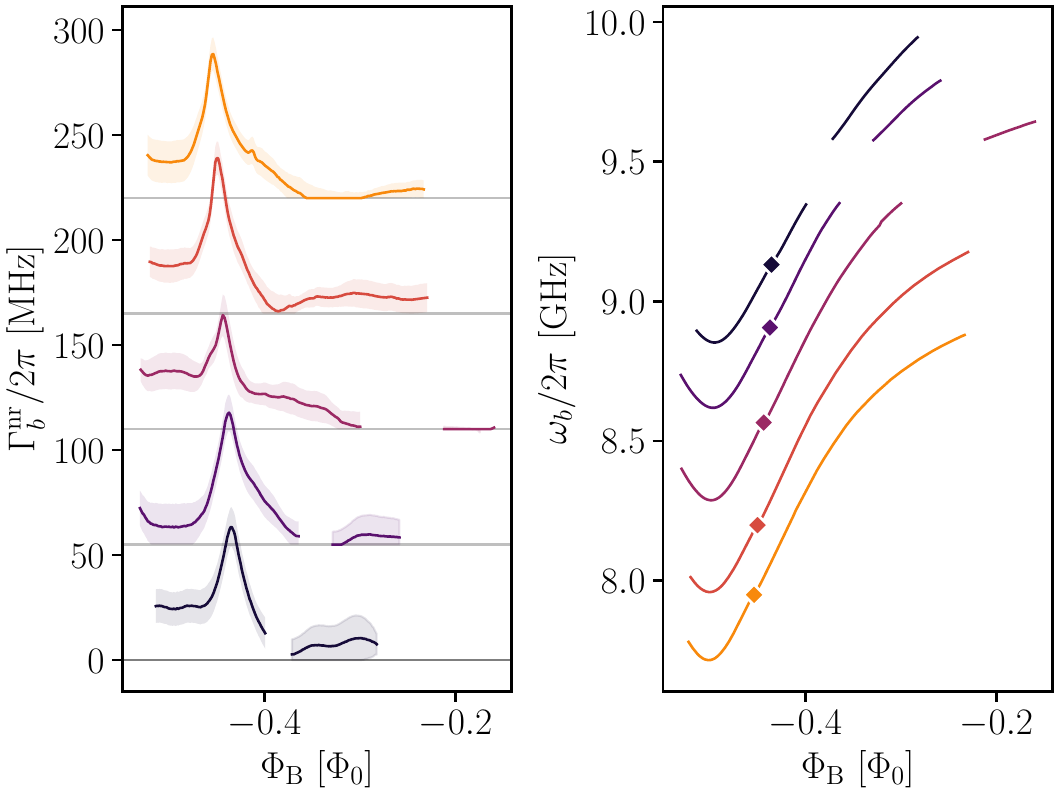}
\caption{\textbf{Left.} Mode $b$ internal losses as a function of the external magnetic flux $\Phi_\mathrm{B}$. Different colours correspond to different array impedances $Z_\mathrm{A}$, ranging from $Z_\mathrm{A}=\qtylist{1100}{\ohm}$ (orange) to $Z_\mathrm{A}=\qtylist{975}{\ohm}$ (black). The curves are offset vertically for clarity. The shaded area around the data shows the uncertainty in the extracted internal losses due to the imperfect isolation of the microwave components used in the output line~\cite{Rieger.20225gr}. For all values of the array impedance, $Z_\mathrm{A}$, the non-radiative width $\Gamma_b^\textrm{nr}$ shows a sharp peak which is attributed to the three-photon down-conversion process. \textbf{Right.} Resonance frequency of mode $b$ as a function of $\Phi_\mathrm{B}$ for different values of $Z_\mathrm{A}$ (same colour code as left panel). The diamond markers show the mode $b$ frequency at the points where the internal losses are maximal (peak on the left panel).}
\label{fig:FigAppB2}
\end{center}
\end{figure}

Tuning the Josephson inductance of the array makes it possible to change its characteristic impedance $Z_\mathrm{A}\left( \Phi_\mathrm{A}\right) = \sqrt{L\left( \Phi_\mathrm{A} \right)/C_\mathrm{g}}$. Figure \ref{fig:FigAppB1} shows that the in-line inductance of the array cannot strictly speaking be considered as a constant within a given period of the $\Phi_\mathrm{B}$ modulation, but the variations in the chain impedance remain small in practice for moderate changes of the flux.
Changing the array impedance also modifies the speed at which waves travel back and forth in the array, resulting in a change in the resonant frequencies of the standing waves. Thus, tuning $\Phi_\textrm{A}$ changes the optimal flux that must be passed through the small junction to achieve the resonance between the states $|3_a,0_b\rangle$ and $|0_a,1_b\rangle$. As a result, the mode $b$ frequency at which conversion is observed is also changed, as indicated by diamonds in the right panel of Figure \ref{fig:FigAppB2}.

This is demonstrated by the left panel of Figure \ref{fig:FigAppB2}, that shows the extracted internal losses of mode $b$ as a function of magnetic flux $\Phi_\mathrm{B}$. The flux range is folded here into five curves as to show five periods of the small junctions, associated each with a different array impedance.
For each period, we observe a sharp increase in the non-radiative losses up to \qtylist{70}{\mega\hertz} above the base level. The losses induced by down-conversion thus greatly exceed extrinsic sources of non-radiative decay, namely $\Gamma^{1\leftrightarrow3} \gtrsim \Gamma_b^{\textrm{nr},0}$, so that the system can be considered to be in a strong down-conversion regime. This observation of resonant losses is confirmed qualitatively by a diagrammatic calculation of the losses according to the theory developed in Ref.~\cite{10.21468/scipostphys.14.5.130}.

\subsection{Power dependence of the fluorescence}
\label{app:power}

\begin{figure}[htb]
\begin{center}
\includegraphics[width = 0.45\textwidth]{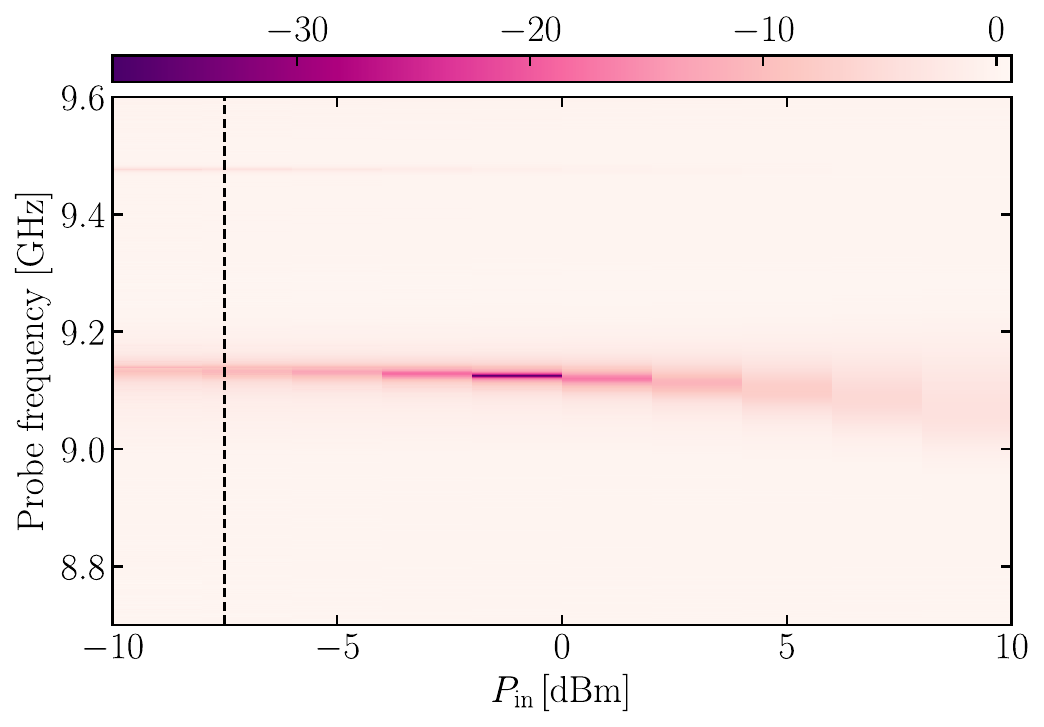}
\caption{Colour map of the magnitude of the reflection coefficient $\vert S_{11}(\omega_\text{d})\vert$ in the vicinity of mode $b$ resonance frequency and as a function of the input power at room temperature. The vertical dashed line shows the power used to perform the multimode fluorescence measurement.}
\label{fig:FigAppC} 
\end{center}
\end{figure}

Choosing the appropriate input power is an important part of measuring the signal from the down-converted photons. On the one hand, it is desirable to maximise the flux of outgoing photons by increasing the input power $P_\textrm{in}$, which clearly improves the signal-to-noise-ratio. On the other hand, one wants to populate mode $b$ ideally with one excitation that will eventually be converted into three excitations of mode $a$ by the photon conversion process. More precisely, the experiment aims at the single-photon regime, defined by an average number of photons in mode $b$ much smaller than one. 

Anharmonicity of the system's modes  allows an intuitive visualisation of the maximum power that can be put in before triggering the nonlinearity of mode $b$. Looking at the power dependence of the reflection coefficient $S_{11}$ plotted in Figure \ref{fig:FigAppC}, one can clearly see the nonlinearity of mode $b$ as the input power is increased. In fact, the resonance frequency shows a clear downward shift as the input power is increased. Since the expected nonlinearity of the modes is comparable to their linewidth, any significant shift in the reflection data suggests an average number of photons in mode $b$ significantly greater than one. We therefore chose to drive mode $b$ with an input power such that no significant shift in the resonance frequency can be resolved, ensuring that the system is in the single photon regime. Unless otherwise stated, the fluorescence spectra presented in the following were measured with an input power of $-7.5 \, \textrm{dBm}$ at room temperature, shown as a vertical dashed line in Figure \ref{fig:FigAppC}. The generated signal is attenuated by $-121.6 \, \textrm{dB} $ before the sample input, resulting in a $-129 \, \textrm{dBm}$ input power referred to the input of the sample.

\subsection{Theoretical calculation of the fluorescence spectrum}
\label{app:spectrum}

\newcommand{\hubbard}{\hat{X}}

\subsubsection{The model}

Our goal is to calculate the fluorescence spectrum to the lowest order in the weak drive strength. For this purpose it is sufficient to keep only five states in the Hilbert space of the two modes: 
%$|0_a0_b\rangle$, $|1_a0_b\rangle$, $|2_a0_b\rangle$, $|3_a0_b\rangle$, $|0_a1_b\rangle$. 
\begin{align}
    &\left\{|0_a0_b\rangle,\:|1_a0_b\rangle,\:|2_a0_b\rangle,\:
    |3_a0_b\rangle,\:|0_a1_b\rangle\right\}\nonumber\\
    {}&{} \equiv \left\{|0\rangle,\,|1\rangle,\,|2\rangle,\,
    |3\rangle,\,|4\rangle\right\}.
    \label{eq:basis}
\end{align}
The two-mode Hamiltonian can be written in terms of the photon creation and annihilation operators of these modes, $\hat{a}^\dagger,\hat{a},\hat{b}^\dagger,\hat{b}$, truncated to the reduced Hilbert space, as
\begin{align}\label{eq:Hab}
    \hat{H}_{ab} = {}&{} \hbar\omega_{10}\,\hat{a}^\dagger\hat{a}
    %+ \frac{\hbar}{2}\,(\omega_{21}-\omega_{10})\,\hat{a}^\dagger\hat{a}^\dagger{}^2\hat{a}^2 \nonumber\\
    %removed one a^dagger: 
    + \frac{\hbar}{2}\,(\omega_{21}-\omega_{10})\,\hat{a}^\dagger{}^2\hat{a}^2 \nonumber\\
    {}&{} + \frac{\hbar}{6}\,(\omega_{32}-2\omega_{21}+\omega_{10})\,\hat{a}^\dagger{}^3\hat{a}^3\nonumber\\
    {}&{}+ \hbar\omega_{b}\hat{b}^\dagger\hat{b}
    + g_{1\leftrightarrow 3}\left( \hat{a}^{\dagger}{}^3 \hat{b}+\hat{b}^{\dagger}\hat{a}^3 \right).
\end{align}
The frequencies $\omega_{10},\omega_{21},\omega_{32}$ of the transitions between different multi-photon states of mode~$\hat{a}$ are treated as independent phenomenological parameters, as discussed in the main text. These three frequencies are assumed to be close to each other (all around 3~GHz), and well detuned from the frequency $\omega_b$ of mode~$\hat{b}$ (around 9~GHz). Thus, we have a hierarchy of energy scales:
\begin{align}\label{eq:hierarchy}
{}&{}    \left|\frac{\omega_b}3-\omega_{10}\right|,\,|\omega_{n+1,n}-\omega_{10}|,\,\frac{g_{1\leftrightarrow 3}}\hbar,\, \Gamma_a,\,\Gamma_b
   \ll \omega_{10},
\end{align}
where $\Gamma_a$, $\Gamma_b$ are the modes' decay rates.

Coupling of the two modes to the external transmission line (TL) is modeled in an almost standard way, with just two nuances. First, to describe the reflection geometry of the experiment, instead of a semi-infinite one-dimensional TL hosting left- and right-travelling waves, we prefer to use an infinite TL with only right travelling waves, whose transmitted wave at $x>0$ thus maps to the reflected wave of the experimental setup (we assume that the modes couple to the $x=0$ point of the TL). Second, since the two modes couple to two well-separated spectral regions of the TL, as guaranteed by the inequality~(\ref{eq:hierarchy}), we couple the two modes to two different effective TLs. Such TLs are described by the Hamiltonian
\begin{align}\label{eq:HTL=}
\hat{H}_\text{TL}=
{}&{}\int_{-\infty}^{+\infty}{dx}\,\hat\alpha^\dagger(x)\left(-i\hbar{v}_\alpha\,\frac{d}{dx}\right)\hat\alpha(x)\nonumber\\
{}&{}+\int_{-\infty}^{+\infty}{dx}\,\hat\beta^\dagger(x)\left(-i\hbar{v}_\beta\,\frac{d}{dx}\right)\hat\beta(x),
\end{align}
where $\hat\alpha(x)$ and $\hat\beta(x)$ are the corresponding field operators, and $v_\alpha,v_\beta$ are the propagation velocities (which may be different if the dispersion of the original physical TL has some curvature). The fact that the Hamiltonian~(\ref{eq:HTL=}) is not bounded from below does not pose any problem for us, since we are interested in the response of a driven system, not in the ground state properties.

The TLs' coupling to the two modes is parametrized by the modes' respective radiative decay rates, $\Gamma_a^\text{r},\Gamma_b^\text{r}$:
\begin{align}
\hat{H}_\text{int} = {}&{}
\sqrt{\hbar^2v_\alpha\Gamma_a^\text{r}} \left[\hat\alpha^\dagger(0)\,\hat{a}+\hat{a}^\dagger\,\hat\alpha(0)\right]
\nonumber\\
{}&{} + \sqrt{\hbar^2v_\beta\Gamma_b^\text{r}} \left[\hat\beta^\dagger(0)\,\hat{b}+\hat{b}^\dagger\,\hat\beta(0)\right].
\end{align}
Then solution of the Heisenberg equation of motion for the $\beta$ TL,
\begin{equation}
\left(\partial_t+v_\beta\partial_x\right)\hat\beta(x,t)=-i\sqrt{v_\beta\Gamma_b^\text{r}}\,\hat{b}(t)\,\delta(x),
\end{equation}
relates the incident and transmitted fields
\begin{equation}\label{eq:input-output}
\hat\beta(x>0,t)=\hat\beta(0^-,t-x/v_\beta)-i\sqrt{\frac{\Gamma_b^\text{r}}{v_\beta}}\,\hat{b}(t-x/v_\beta),
\end{equation}
and similarly for the $\alpha$ TL and the $a$~mode. We denote $\hat\alpha(0^-,t)\equiv\hat\alpha_\mathrm{in}(t)$, $\hat\beta(0^-,t)\equiv\hat\beta_\mathrm{in}(t)$, the input fields.

Non-radiative decay of photons in the two modes can be included by coupling them to additional dissipative baths. The corresponding Hamiltonian can be taken in the same form as for the TL. We do not write it explicitly to avoid cluttering the text; the non-radiative decay rates $\Gamma_a^\text{nr},\Gamma_b^\text{nr}$ will be introduced straightforwardly in the final expressions.

The classical monochromatic drive with frequency $\omega_\text{d}$ can be included either as a term in the Hamiltonian,
\begin{equation}\label{eq:Hd=}
    \hat{H}_\text{d} = V_\text{d}^*e^{i\omega_\text{d}t}\hat{b} + V_\text{d}e^{-i\omega_\text{d}t}\hat{b}^\dagger,
\end{equation}
or as a coherent signal in the TL which amounts to a shift of the Heisenberg field operator, 
\begin{equation}
\hat\beta(x,t) \to \hat\beta(x,t) + \frac{V_\text{d}e^{i\omega_\text{d}(x/v_\beta-t)}}{\sqrt{\hbar^2v_\beta\Gamma_b^\text{r}}},
\end{equation}
which we will refer to as the driven mode and the driven TL picture, respectively. The latter picture is useful to define the reflection coefficient of the incident beam as $S_{11}(\omega_\text{d})=\langle\beta(0^+,t)\rangle_c/\langle\beta(0^-,t)\rangle_c$ for the field operators before the shift (that is, both averages are taken over the coherent state of the incident field, they have the same monochromatic time dependence, which thus cancels out). It can be equivalently written as
\begin{equation}\label{eq:reflection}
S_{11}(\omega_\text{d}) =1
-i\,\frac{\Gamma_b^\text{r}\,\langle\hat{b}(t)\rangle}{V_\mathrm{d}e^{-i\omega_\text{d}t}},
\end{equation}
where the average is taken over the TL vacuum state and the dynamics of the Heisenberg operator $\hat{b}(t)$ is determined including the drive Hamiltonian~(\ref{eq:Hd=}). 
The driven TL picture also enables us to relate the drive amplitude $V_\text{d}$ to the input power $J_\text{in}$, or the energy current at $x<0$ (given by the TL Hamiltonian density multiplied by the velocity~$v_\beta$, and averaged over the coherent state of the incident field):
\begin{align}
    J_\text{in} = \left\langle{v}_\beta\,\hat\beta^\dagger(x,t)\left(-i\hbar{v}_\beta\,\partial_x\right)\hat\beta(x,t)\right\rangle_c = \frac{\omega_\text{d}|V_\text{d}|^2}{\hbar\Gamma_b^\text{r}}.
    \label{eq:input_power}
\end{align}
Having justified Eqs.~(\ref{eq:reflection}) and~(\ref{eq:input_power}), we will work in the driven mode picture from now on.

To deduce the expression for the fluorescence spectrum, we note that the time-averaged energy current $\overline{J}$ in the $\alpha$~TL at some point $x>0$ (time averaging is necessary since the system is driven by a time-dependent Hamiltonian~$\hat{H}_\text{d}$),
\begin{align}
    \overline{J} = {}&{} \lim_{T\to\infty}\int_0^T\frac{dt}{T}\,\langle\hat{J}(x,t)\rangle\nonumber\\
    = {}&{} \lim_{T\to\infty}\int_0^T\frac{dt}{T}\left\langle{v}_\alpha\,\hat\alpha^\dagger(x,t)\left(-i\hbar{v}_\alpha\,\partial_x\right)\hat\alpha(x,t)\right\rangle,
\end{align}
can be represented in terms of its spectral density,
\begin{subequations}\begin{align}
    & \overline{J} = \int_{-\infty}^\infty\frac{d\omega}{2\pi}
    \,e^{-i\omega{t}}\,J_\omega,\\
    &J_\omega = v_\alpha\hbar\omega
    \int\limits_{-\infty}^\infty{d}\tau\,e^{i\omega\tau}
    \lim\limits_{T\to\infty}\int_0^T\frac{dt}{T}\,\langle\hat\alpha^\dagger(x,t)\,\hat\alpha(x,t+\tau)\rangle
    \nonumber\\
    &\quad {} = {}\hbar\omega\Gamma_a^\text{r}
    \int\limits_{-\infty}^\infty{d}\tau\,e^{i\omega\tau}
    \lim\limits_{T\to\infty}\int_0^T\frac{dt}{T}\,\langle\hat{a}^\dagger(t)\,\hat{a}(t+\tau)\rangle\nonumber\\
    &\quad {}\equiv\hbar\omega{I}_\omega,
    \label{eq:spectrum_correlator}
\end{align}\end{subequations}
where $I_\omega$ is the spectral density of the emitted photons (since each photon carries the energy $\hbar\omega$).
Thus, we have to calculate the two-time correlator $\langle\hat{a}^\dagger(t)\,\hat{a}(t+\tau)\rangle$ for a driven few-level system, coupled to a dissipative bath in the vacuum state. This is the standard object that arises in a calculation of the fluorescence spectrum, and the standard way to calculate it is the quantum regression theorem~\cite{Mollow.1969}.

\subsubsection{Equations of motion}

First of all, we get rid of the time dependence in the Hamiltonian by passing to the rotating frame. Namely, we apply a time-dependent unitary transformation to the full system's wave function $|\Psi(t)\rangle=\hat{U}(t)|\tilde\Psi(t)\rangle$ with $\hat{U}(t)=e^{-i\hat{\mathcal{N}}\omega_\mathrm{d}{t}}$, where $\hat{\mathcal{N}}$ is the total number of excitations conserved by the Hamiltonian $\hat{H}_{ab}+\hat{H}_\text{TL}+\hat{H}_\text{int}$:
\begin{equation}
\hat{\mathcal{N}}\equiv\frac13\,\hat{a}^\dagger\hat{a}+\hat{b}^\dagger\hat{b}
+\int\limits_{-\infty}^\infty{dx}\left[\frac13\,\hat\alpha^\dagger(x)\,\hat\alpha(x)+\hat\beta^\dagger(x)\,\hat\beta(x)\right].
\end{equation}
Then, if the original wave function $|\Psi(t)\rangle$ satisfies the Schr\"odinger equation with the full Hamiltonian $\hat{H}=\hat{H}_{ab}+\hat{H}_\text{TL}+\hat{H}_\text{int}+\hat{H}_\text{d}$, the transformed wave function~$|\tilde\Psi(t)\rangle$ satisfies the Schr\"odinger equation with the transformed Hamiltonian $\hat{\tilde{H}}=\hat{U}^\dagger\hat{H}\hat{U}-i\hbar\hat{U}^\dagger(d\hat{U}/dt)$. Since $\hat{U}^\dagger\hat{a}\hat{U}=\hat{a}e^{-i\omega_\mathrm{d}t/3}$, $\hat{U}^\dagger\hat{b}\hat{U}=\hat{b}e^{-i\omega_\mathrm{d}t}$, $\hat{\tilde{H}}$ is obtained from~$\hat{H}$ by removing $e^{\pm{i}\omega_\mathrm{d}t}$ in the drive term $\hat{H}_\text{d}$, and shifting the frequencies
\begin{align}
    \tilde\omega_{n+1,n}=\omega_{n+1,n}-\frac{\omega_\text{d}}{3},\quad
    \tilde\omega_b=\omega_b-\omega_\text{d}.
\end{align}
From now on, we will work in the rotating frame, so we will suppress the tildes over all operators (still, keeping them at the frequencies, to keep the relation to the original parameters).

Next, we introduce the Hubbard operator basis $\hubbard_{jj'}\equiv|j\rangle\langle{j}'|$, where the states $|j\rangle$ with $j=0,\ldots,4$ are defined in Eq.~(\ref{eq:basis}). Then, the average of the Heisenberg operator $\langle\hubbard_{jj'}(t)\rangle=\rho_{j'j}(t)$, the corresponding matrix element of the Schr\"odinger picture density matrix. We have the obvious algebra $\hubbard_{jj'}\hubbard_{j''j'''}=\delta_{j'j''}\hubbard_{jj'''}$, and $\hat{a}=\hubbard_{01}+\sqrt{2}\hubbard_{12}+\sqrt{3}\hubbard_{23}$, $\hat{b}=\hubbard_{04}$. The modes' Hamiltonian can be written as
\begin{align}
    \hat{{H}}_{ab} + \hat{{H}}_\text{d} = {}&{}
    \hbar\tilde\omega_{10}\hubbard_{11}
    + \hbar(\tilde\omega_{21}+\tilde\omega_{10})\hubbard_{22}
    \nonumber\\
    {}&{}+\hbar(\tilde\omega_{32}+\tilde\omega_{21}+\tilde\omega_{10})\hubbard_{33}
    + \hbar\tilde\omega_b\hubbard_{44}\nonumber\\
    {}&{} +
    \sqrt6g_{1\leftrightarrow 3}(\hubbard_{34}+\hubbard_{43})
    + \hat{V}_\text{d}^*\hubbard_{04} + \hat{V}_\text{d}\hubbard_{40}.
\end{align}
Next, we write the Heisenberg equations of motion for the operators $\hubbard_{nn'}$. Commutation with $\hat{{H}}_\text{int}$ introduces the TL field operators, which are found from the ready solution~(\ref{eq:input-output}). When this solution is plugged back into the equations for $\hubbard_{jj'}$, we regularized the zero argument Heaviside function as %$\theta(0)=1/2$.
\begin{align}
    \theta(0) = {}&{} \int_{-\infty}^\infty{d}x\,\delta(x)\,\theta(x) = \int_{-\infty}^\infty{d}x\,\delta(x)\int_{-\infty}^x{d}x'\,\delta(x')\nonumber\\
    = {}&{} \int_0^1\theta\,d\theta=\frac12,
\end{align}
where $\theta(x)$ and $\delta(x)$ are the Heaviside and Dirac functions, which can be regularized by smearing them over a small range of~$x$ [the argument $x=0$ in $\hat{H}_\text{int}$ can be understood as a convolution with the regularized $\delta(x)$].

We then separate the full set of equations of motion for $\hubbard_{nn'}$ into two subsets:
\begin{widetext}
\begin{subequations}\label{eq:Lindblad-offdiag}\begin{align}
\frac{d\hubbard_{31}}{dt} {}&{} =(i\tilde\omega_{31} - 2\Gamma_a^\text{r})\hubbard_{31}
%+ \frac{i}{\hbar}\,\sqrt{6}g_{1\leftrightarrow3}
+ i\chi\hubbard_{41}
+ i\sqrt{v_\alpha\Gamma_a^\text{r}}\hat\alpha_\mathrm{in}^\dagger\left(\sqrt{3}\hubbard_{21}-\sqrt{2}\hubbard_{32}\right)
- i\sqrt{v_\alpha\Gamma_a^\text{r}}\hubbard_{30}\hat\alpha_\mathrm{in},\\
\frac{d\hubbard_{41}}{dt} {}&{} =\left(i\tilde\omega_b-i\tilde\omega_{10}
-\frac{\Gamma_a^\text{r}+\Gamma_b^\text{r}}{2}\right)\hubbard_{41}
%+ \frac{i}{\hbar}\,\sqrt{6}g_{1\leftrightarrow3}
+ i\chi\hubbard_{31}
+ \frac{iV_\text{d}^*}\hbar\,\hubbard_{01}
+ i\sqrt{v_\beta\Gamma_b^\text{r}}\hat\beta^\dagger_\mathrm{in}\hubbard_{01}
- i\sqrt{v_\alpha\Gamma_a^\text{r}}\left(\hubbard_{40}\hat\alpha_\mathrm{in}
+\hat\alpha_\mathrm{in}^\dagger\sqrt{2}\hubbard_{42}\right),\\
\frac{d\hubbard_{01}}{dt} {}&{} = \left({-i}\tilde\omega_{10}-\frac{\Gamma_a^\text{r}}2\right)\hubbard_{01}
+ \frac{iV_\text{d}}\hbar\,\hubbard_{41}
+\Gamma_a^\text{r}\sqrt{2}\hubbard_{12}
+i\sqrt{v_\alpha\Gamma_a^\text{r}}\left(\hubbard_{11}-\hubbard_{00}\right)\hat\alpha_\mathrm{in}
-i\sqrt{v_\alpha\Gamma_a^\text{r}}\hat\alpha^\dagger_\mathrm{in}\sqrt{2}\hubbard_{02},\\
\frac{d\hubbard_{12}}{dt} {}&{} = \left({-i}\tilde\omega_{21}-\frac{3\Gamma_a^\text{r}}2\right)\hubbard_{12}
+\Gamma_a^\text{r}\sqrt{6}\hubbard_{23}
+i\sqrt{v_\alpha\Gamma_a^\text{r}}\left(\sqrt{2}\hubbard_{22}-\sqrt{2}\hubbard_{11}\right)\hat\alpha_\mathrm{in}
+i\sqrt{v_\alpha\Gamma_a^\text{r}}\hat\alpha^\dagger_\mathrm{in}\left(\hubbard_{02}-\sqrt{3}\hubbard_{13}\right),\\
\frac{d\hubbard_{23}}{dt}{}&{} = \left({-i}\tilde\omega_{32}-\frac{5\Gamma_a^\text{r}}2\right)\hubbard_{23}
%-\frac{i}{\hbar}\,\sqrt{6}g_{1\leftrightarrow3}
-i\chi\hubbard_{24}
+i\sqrt{v_\alpha\Gamma_a^\text{r}}\left(\sqrt{3}\hubbard_{33}-\sqrt{3}\hubbard_{22}\right)\hat\alpha_\mathrm{in}
+i\sqrt{v_\alpha\Gamma_a^\text{r}}\hat\alpha^\dagger_\mathrm{in}\sqrt{2}\hubbard_{13},\\
\frac{d\hubbard_{24}}{dt} {}&{} =\left(i\tilde\omega_{20}-i\tilde\omega_b-\Gamma_a^\text{r}-\frac{\Gamma_b^\text{r}}{2}\right)\hubbard_{24}
%- \frac{i}{\hbar}\,\sqrt{6}g_{1\leftrightarrow3}
-i\chi\hubbard_{23}
- \frac{iV_\text{d}}\hbar\,\hubbard_{20}
- i\sqrt{v_\beta\Gamma_b^\text{r}}\hubbard_{20}\hat\beta_\mathrm{in}
+ i\sqrt{v_\alpha\Gamma_a^\text{r}}\left(\sqrt{3}\hubbard_{34}\hat\alpha_\mathrm{in}
+ \hat\alpha_\mathrm{in}^\dagger\sqrt{2}\hubbard_{14}\right),\\
%
%added a missing term
\frac{d\hubbard_{20}}{dt} {}&{} = \left(i\tilde\omega_{20}-\Gamma_a^\text{r}\right)\hubbard_{20}
+ \Gamma_a^\text{r}\sqrt{3}\hubbard_{31}
- \frac{iV_\text{d}^*}\hbar\,\hubbard_{24}
+ i\sqrt{v_\alpha\Gamma_a^\text{r}}\hat\alpha_\mathrm{in}^\dagger\left(\sqrt{2}\hubbard_{10} - \hubbard_{21}\right)
+ i\sqrt{v_\alpha\Gamma_a^\text{r}}\sqrt{3}\hubbard_{30}\hat\alpha_\mathrm{in},
\end{align}\end{subequations}
and
\begin{subequations}\label{eq:Lindblad-diag}\begin{align}
    \frac{d\hubbard_{04}}{dt} {}&{} =\left(-i\tilde\omega_b-\frac{\Gamma_b^\text{r}}{2}\right)\hubbard_{04} 
    %- \frac{i}{\hbar}\,\sqrt{6}g_{1\leftrightarrow3}
    -i\chi\hubbard_{03}
    +\frac{iV_\text{d}}{\hbar}\left(\hubbard_{44}-\hubbard_{00}\right)
    +i\sqrt{v_\beta\Gamma_b^\text{r}}\left(\hubbard_{44}-\hubbard_{00}\right)\hat\beta_\text{in},\\
    \frac{d\hubbard_{03}}{dt} {}&{} =\left(-i\tilde\omega_{30}-\frac{3\Gamma_a^\text{r}}{2}\right)\hubbard_{03} 
    %- \frac{i}{\hbar}\,\sqrt{6}g_{1\leftrightarrow3}
    -i\chi\hubbard_{04} + \frac{iV_\text{d}}\hbar\,\hubbard_{43} 
    +i\sqrt{v_\alpha\Gamma_a^\text{r}}\left(\hubbard_{13}-\sqrt{3}\hubbard_{02}\right)\hat\alpha_\text{in},\\
    \frac{d\hubbard_{34}}{dt} {}&{} = \left(i\tilde\omega_{30}-i\tilde\omega_b-\frac{3\Gamma_a^\text{r}+\Gamma_b^\text{r}}{2}\right)\hat{X}_{34}
    + i\chi(\hubbard_{44}-\hubbard_{33})
    - \frac{iV_\text{d}}\hbar\,\hubbard_{30} 
    + i\sqrt{v_\alpha\Gamma_a^\text{r}}\hat\alpha_\text{in}^\dagger\hubbard_{24}
    - i\sqrt{v_\beta\Gamma_b^\text{r}}\hubbard_{30}\hat\beta_\text{in},\\
    \frac{d\hubbard_{44}}{dt} {}&{} = -\Gamma_b^\text{r}\hubbard_{44}
    %+ \frac{i}{\hbar}\,\sqrt{6}g_{1\leftrightarrow3}
    + i\chi\left(\hubbard_{34}-\hubbard_{43}\right)
    +\frac{i}{\hbar}\left(V_\text{d}^*\hubbard_{04}-V_\text{d}\hubbard_{40}\right)
    + i\sqrt{v_\beta\Gamma_b^\text{r}}\left(\hat\beta_\text{in}^\dagger\hubbard_{04}-\hubbard_{40}\hat\beta_\text{in}\right),\\
    \frac{d\hubbard_{33}}{dt}{}&{} = -3\Gamma_a^\text{r}\hubbard_{33}
    %-\frac{i}{\hbar}\,\sqrt{6}g_{1\leftrightarrow3}
    -i\chi\left(\hubbard_{34}-\hubbard_{43}\right)
    -i\sqrt{v_\alpha\Gamma_a^\text{r}}\sqrt{3}\hubbard_{32}\hat\alpha_\mathrm{in}
    +i\sqrt{v_\alpha\Gamma_a^\text{r}}\hat\alpha_\mathrm{in}^\dagger\sqrt{3}\hubbard_{23},\\
    \frac{d\hubbard_{22}}{dt} {}&{} = \Gamma_a^\text{r}\left(3\hubbard_{33}-2\hubbard_{22}\right)
    +i\sqrt{v_\alpha\Gamma_a^\text{r}}\left(\sqrt{3}\hubbard_{32}-\sqrt{2}\hubbard_{21}\right)\hat\alpha_\mathrm{in}
    -i\sqrt{v_\alpha\Gamma_a^\text{r}}\hat\alpha_\mathrm{in}^\dagger\left(\sqrt{3}\hubbard_{23}-\sqrt{2}\hubbard_{12}\right),\\
    \frac{d\hubbard_{11}}{dt} {}&{} = \Gamma_a^\text{r}\left(2\hubbard_{22}-\hubbard_{11}\right)
    +i\sqrt{v_\alpha\Gamma_a^\text{r}}\left(\sqrt{2}\hubbard_{21}-\hubbard_{10}\right)\hat\alpha_\mathrm{in}
    -i\sqrt{v_\alpha\Gamma_a^\text{r}}\hat\alpha_\mathrm{in}^\dagger\left(\sqrt{2}\hubbard_{12}-\hubbard_{01}\right),\\
    %
    %\frac{d\hubbard_{00}}{dt} {}&{} = \Gamma_b^\text{r}\hubbard_{11}
    %- \frac{i}{\hbar}\left(V_\text{d}^*\hubbard_{04}-V_\text{d}\hubbard_{40}\right)
    %+i\sqrt{v_\alpha\Gamma_a^\text{r}}\hubbard_{10}\hat\alpha_\mathrm{in}-%i\sqrt{v_\alpha\Gamma_a^\text{r}}\hat\alpha_\mathrm{in}^\dagger\hubbard_{01}.
    %fixed the X_11 term and added the X_44 term
    \frac{d\hubbard_{00}}{dt} {}&{} = \Gamma_a^\text{r}\hubbard_{11}
    + \Gamma_b^\text{r}\hubbard_{44}
    - \frac{i}{\hbar}\left(V_\text{d}^*\hubbard_{04}-V_\text{d}\hubbard_{40}\right)
    +i\sqrt{v_\alpha\Gamma_a^\text{r}}\hubbard_{10}\hat\alpha_\mathrm{in}-i\sqrt{v_\alpha\Gamma_a^\text{r}}\hat\alpha_\mathrm{in}^\dagger\hubbard_{01},
\end{align}\end{subequations}
where we denoted $\tilde\omega_{30}\equiv\tilde\omega_{32} + \tilde\omega_{21} +\tilde\omega_{10}$, $\tilde\omega_{31}\equiv\tilde\omega_{32} + \tilde\omega_{21}$, $\tilde\omega_{20}\equiv \tilde\omega_{21} +\tilde\omega_{10}$, $\sqrt{6}g_{1\leftrightarrow3}/\hbar\equiv\chi$ for compactness.
Upon averaging over the vacuum state, all terms containing the input fields vanish, since $\hat\alpha_\text{in},\hat\beta_\text{in}$ annihilate the vacuum. 
The resulting equations for $\langle\hubbard_{jj'}\rangle$ are nothing but the components of the Lindblad equation for the driven 5-level system, coupled to a bath.  
Then each subset of equations, the averaged Eqs.~(\ref{eq:Lindblad-offdiag}) and Eqs.~(\ref{eq:Lindblad-diag}), is closed.

Labeling the components of $\langle\hubbard_{jj'}\rangle$ by a single index $\mu\in\{31,\,41,\,01,\,12,\,23,\,24,\,20\}$, we can write the averaged Eqs.~(\ref{eq:Lindblad-offdiag}) as
\begin{equation}\label{eq:master-equation}
\frac{d\langle\hubbard_\mu\rangle}{dt}=-\sum_{\mu'}\mathcal{M}_{\mu\mu'}\langle\hubbard_{\mu'}\rangle,
\end{equation}
with a matrix $\mathcal{M}$ given by
{\small
\begin{equation}
\mathcal{M}=\begin{pmatrix}
2\Gamma_a - i\tilde\omega_{31} & -i\chi & 0 & 0 & 0 & 0 & 0 \\
-i\chi & (\Gamma_a+\Gamma_b)/2+i\tilde\omega_{10}-i\tilde\omega_b
& -(i/\hbar)V_\text{d}^* & 0 & 0 & 0 & 0 \\
0 & -(i/\hbar)V_\text{d} & \Gamma_a/2+i\tilde\omega_{10} & -\sqrt{2}\Gamma_a & 0 & 0 & 0 \\
0 & 0 & 0 & 3\Gamma_a/2+i\tilde\omega_{21} & -\sqrt{6}\Gamma_a & 0 & 0 \\
0 & 0 & 0 & 0 & 5\Gamma_a/2+i\tilde\omega_{32} & i\chi & 0 \\
0 & 0 & 0 & 0 & i\chi & \Gamma_a+\Gamma_b/2+i\tilde\omega_b-i\tilde\omega_{20} & iV_\text{d}/\hbar \\
%added matrix element 7,1
-\sqrt{3}\Gamma_a & 0 & 0 & 0 & 0 & iV_\text{d}^*/\hbar & \Gamma_a-i\tilde\omega_{20}
\end{pmatrix}.
\end{equation}
}%
Note that we replaced the radiative decay rates $\Gamma_{a,b}^\text{r}$ by the full ones $\Gamma_{a,b}=\Gamma_{a,b}^\text{r}+\Gamma_{a,b}^\text{nr}$. Indeed, it can be seen straightforwardly that coupling to additional dissipative baths would just produce additional decay terms in the equations of motion. 

\subsubsection{Quantum regression theorem}

The stationary solution of Eq.~(\ref{eq:master-equation}) is $\langle\hubbard_\mu\rangle=0$, implying $\langle\hat{a}\rangle=0$, so there is no coherent radiation at frequencies near $\omega_\mathrm{d}/3$.
The incoherent spectrum~(\ref{eq:spectrum_correlator}) can be found using quantum regression theorem which states that the set of correlators $\langle\hat{O}(0)\,\hubbard_\mu(t>0)\rangle$ with an arbitrary operator $\hat{O}$ satisfies the same equations as $\langle\hubbard_\mu(t)\rangle$, but  with $\langle\hat{O}\hubbard_\mu\rangle$ as initial conditions. Introducing the coefficients $a_\mu$ such  that $\hat{a}=\sum_\mu{a}_\mu\hubbard_\mu$ (that is, $a_{01}=1$, $a_{12}=\sqrt{2}$, $a_{23}=\sqrt{3}$, the rest of $a_\mu$ being zero), we can write
\begin{align}
\langle\hat{a}^\dagger(0)\,\hat{a}(t>0)\rangle=\sum_{\mu,\mu',\mu''}a_\mu\left(e^{-\mathcal{M}t}\right)_{\mu\mu'}a_{\mu''}^*\langle\hubbard_{\mu''}^\dagger\hubbard_{\mu'}\rangle.
\end{align}
At the same time,
$\langle\hat{a}^\dagger(0)\,\hat{a}(t<0)\rangle
=\langle\hat{a}^\dagger(-t)\,\hat{a}(0)\rangle
=\langle\hat{a}^\dagger(0)\,\hat{a}(-t)\rangle^*$,
so the spectrum
\begin{align}
&\int_{-\infty}^\infty
\langle\hat{a}^\dagger(0)\,\hat{a}(t)\rangle{e}^{i\tilde\omega{t}}\,dt
%{}&{} =\int\limits_0^\infty{d}t\,{e}^{i\tilde\omega{t}}\sum_{\mu,\mu'}b_\mu\left(e^{-\mathcal{M}t}\right)_{\mu\mu'}\langle\hat{b}^\dagger\hubbard_{\mu'}\rangle +\mbox{c.c.} \nonumber\\
 = \sum_{\mu,\mu',\mu''}a_\mu(\mathcal{M}-i\tilde\omega)^{-1}_{\mu\mu'}{\langle\hubbard_{\mu'}^\dagger\hubbard_{\mu''}\rangle}^* a_{\mu''}^*+\mbox{c.c.},
 \label{eq:adaggera=}
\end{align}
where we denoted $\tilde\omega\equiv\omega-\omega_\mathrm{d}/3$ and used the fact that all eigenvalues of $\mathcal{M}$ have positive real parts.
The averages ${\langle\hubbard_{j_1j_2}^\dagger\hubbard_{j_3j_4}\rangle}^*=\langle\hubbard_{j_4j_2}\rangle$ in this expression are those determined by the averaged Eqs.~(\ref{eq:Lindblad-diag}). 
\begin{comment}
\begin{equation}
    \begin{pmatrix}
    0 \\ 0 \\ \langle\hubbard_{11}\rangle \\ \sqrt{2}\langle\hubbard_{22}\rangle \\
    \sqrt{3} \langle\hubbard_{33}\rangle \\ \sqrt{3}\langle\hubbard_{34}\rangle \\ \sqrt{3}\langle\hubbard_{30}\rangle
    \end{pmatrix}
\end{equation}
\end{comment}
To the leading order in the drive, we find $\langle\hubbard_{00}\rangle=1+O(V_\text{d}^2)$, $\langle\hubbard_{11}\rangle=2\langle\hubbard_{22}\rangle=3\langle\hubbard_{33}\rangle$, while for $j,j'=3,4$ we can write $\langle\hubbard_{j0}\rangle=\psi_j^*$, $\langle\hubbard_{jj'}\rangle=\psi_j^*\psi_{j'}$, with 
\begin{align}
\begin{pmatrix} \psi_3 \\ \psi_4 \end{pmatrix} 
= {}&{} \frac{-i{V}_\text{d}/\hbar}{(\Gamma_b/2+i\tilde\omega_b)(3\Gamma_a/2+i\tilde\omega_{30}) + \chi^2}
\begin{pmatrix}
    -i\chi \\ 3\Gamma_a/2+i\tilde\omega_{30}
\end{pmatrix}.
\end{align}
As a result, Eq.~(\ref{eq:spectrum_correlator}) evaluates to an analytical expression
\begin{align}
    I_\omega{}&{} = 6\Gamma_a^\text{r}|\psi_3|^2\,\Re
    \frac{3(3\Gamma_a/2-i\tilde\omega)^2 + 3(\Gamma_a/2)(i\tilde\omega_{10}-i\tilde\omega_{21})+ \tilde\omega_{10}^2+ \tilde\omega_{10} \tilde\omega_{21}+ \tilde\omega_{21}^2}%
{[\Gamma_a/2+i (\tilde\omega_{10}-\tilde\omega)] [3 \Gamma_a/2+i(\tilde\omega_{21}-\tilde\omega)]
[\gamma_a-i (\tilde\omega_{20}+\tilde\omega)]}.
\label{eq:spectrum_analytical}
\end{align}
\end{widetext}
Eq.~(\ref{eq:spectrum_analytical}) was used to generate the theoretical emission spectrum, shown in the main text.
It has three poles corresponding to three emission peaks centered at $\omega=\omega_{10},\,\omega_{21},\,\omega_\text{d}-(\omega_{10}+\omega_{21})$, respectively. The first two are straightforwardly understood as emitted in the transitions $|2\rangle\to|1\rangle\to|0\rangle$, with real populations of the states $|1\rangle$ and $|2\rangle$. The third frequency is determined by the energy conservation: the energies of the three emitted photons must add up exactly to $\hbar\omega_\text{d}$, the energy of the incident photon. Thus, the states $|3\rangle$ and $|4\rangle$ are populated virtually, since the energy would not be conserved otherwise. However, this restriction is relaxed if the system is allowed to emit low-frequency excitations without performing transitions between different states~$|n\rangle$; such processes are known as pure dephasing.

\subsection{The role of pure dephasing}
\label{app:dephasing}
Here we investigate dephasing, which we have ignored in the main text.
Dephasing is caused by the coupling of the modes via $\hat{a}^\dagger\hat{a}$ and $\hat{b}^\dagger\hat{b}$ to some external systems whose dynamics are slow compared to that of the device. We therefore model dephasing by adding stochastic terms
$\xi_a(t)\, \hat a^\dagger \hat a$ and $\xi_b(t)\, \hat b^\dagger \hat b$ to the Hamiltonian, and averaging final results over 
$\xi_a(t)$ and $\xi_b(t)$. For simplicity, we assume Gaussian white noise. 
There are two alternatives: if the dominant source of dephasing is flux noise, $\xi_a(t)$ and $\xi_b(t)$ will be correlated:
\begin{equation}
\overline{\xi_j(t)\xi_{j'}(t')}=\sqrt{\gamma_j\gamma_{j'}} \delta(t-t'),~j,\,{j'}\in\{a,b\}.\label{dephase:2}
\end{equation}
On the other hand, if modes $a$ and $b$ couple to different external systems, $\xi_a(t)$ and $\xi_b(t)$ will be uncorrelated:
\begin{equation}
\overline{\xi_j(t)\xi_{j'}(t')}=\delta_{jj'}\gamma_j^2 \delta(t-t'),~j,\,j'\in\{a,b\}.\label{dephase:1}
\end{equation}
We explicitly show here the analysis corresponding to the former case (\ref{dephase:2}), and indicate the small modifications
to results corresponding to case (\ref{dephase:1})  

The terms $\xi_a(t) \hat a^\dagger \hat a$ and $\xi_b(t) \hat b^\dagger \hat b$  give a random phase to off-diagonal density matrix elements (in the Hubbard basis).
When we average over noise, this leads to a decay of these matrix elements. In Sec.\,\ref{app:spectrum} it was found that without dephasing, states
$\left|3\right>$ and $\left|4\right>$, that respectively contain 3 photons in mode $a$ and 1 photon in mode $b$, are only populated
virtually. The situation changes when dephasing induces the decay of off-diagonal density matrix elements. As a result, the spectral density of down-converted photons
acquires two further poles.
One is at $\omega_{32}$. It corresponds to a process that starts with three photons
in mode $a$. The other is at $\omega_b-\omega_{20}$, and corresponds to a process that starts with one photon in mode~$b$.
We investigate how these additional poles affect the theoretical prediction for the down-conversion spectrum.

With $\mu=(n_a,n_b,n_a',n_b')$, and 
\begin{equation}
\hat \rho_\mu=\frac{(\hat{b}^\dagger)^{n_b}(\hat{a}^\dagger)^{n_a}\left|0\right>\left<0\right|\hat{a}^{n_a'}\hat{b}^{n_b'}}{\sqrt{n_a!\,n_b!\,n_a'!\,n_b'!}},
\end{equation}
we have that
\begin{align}
&\left[\xi_a(t)\,\hat{a}^\dagger \hat{a}+\xi_b(t)\,\hat{b}^\dagger\hat{b},\,\hat \rho_\mu\right]\nonumber\\
&~~~~~=\left[\xi_a(t)(n_a-n_a')+\xi_b(t)(n_b-n_b')\right]\hat \rho_\mu.
\end{align}
We collect the averages $ \left<\hat\rho_\mu(t)\right>$ of the Heisenberg operators $\hat\rho_\mu(t)$ into a vector $\bm \rho(t)$. 
If the master equation for~$\bm\rho$ reads $d\bm\rho/dt=\mathcal L \bm \rho$, then, with noise included, we have
\begin{equation}
\frac{d\bm\rho}{dt}=\left[\mathcal L + i \Xi(t)\mathcal D\right]\bm \rho,
\end{equation}
where the superoperator $\mathcal D$ is diagonal:
\begin{equation}
[\mathcal D]_{\mu\nu}=\delta_{\mu\nu}\left[\sqrt{\gamma_a}(n_a-n_a')+\sqrt{\gamma_b}(n_b-n_b')\right],
\end{equation}
and $\overline{\Xi(t)\Xi(t')}=\delta(t-t')$. 
We will derive a master equation for the noise-averaged $\overline{\bm \rho(t)}$. Define
\begin{equation}
\bm \rho_\text{int}(t)=e^{-\mathcal L t}\bm \rho(t),
\end{equation}
so that
\begin{equation}
\frac{d}{dt}\bm \rho_\text{int}(t)=i\Xi(t)\mathcal D_\text{int}(t)\bm \rho_\text{int}(t),~~~\mathcal D_\text{int}(t)=e^{-\mathcal L t}\mathcal D e^{\mathcal L t}.
\end{equation}
The solution involves a time-ordered exponential. For $t>t'$,
\begin{equation}
\bm \rho_\text{int}(t)=\text{Texp}\left[\,i\int_{t'}^t dt'' \Xi(t'') \mathcal D_\text{int}(t'')\right]\,\bm\rho_\text{int}(t').
\end{equation}
To average over noise, we discretize time as $t_\alpha=t' + \alpha \Delta t$, with $\Delta t=(t-t')/N$, and approximate
\begin{equation}
\text{Texp}\left[i\int_{t_\alpha}^{t_{\alpha+1}} dt'' \Xi(t'') \mathcal D_\text{int}(t'')\right]\simeq \exp\left[ i \Delta t\, \Xi_\alpha \mathcal D_\text{int}(t_\alpha)\right],
\end{equation}
where $\Xi_\alpha$ are discrete Gaussian variables with $\overline{\Xi_\alpha \Xi_{\alpha'}}=\delta_{\alpha\alpha'}/\Delta t$. We recover the exact result after sending $N\to\infty$.
Gaussian averaging gives
\begin{equation}
\overline{\exp\left[ i \Delta t \Xi_\alpha \mathcal D_\text{int}(t_\alpha)\right]}= \exp\left[ -\Delta{t}\,\mathcal D_\text{int}(t_\alpha)^2/2\right], 
\end{equation}
and hence
\begin{equation}
\overline{\bm \rho(t)}=\lim_{N\to\infty}\left[e^{- \mathcal D^2 \Delta t/2}e^{\mathcal L \Delta t}\right]^N \overline{\bm \rho(t')}.
\end{equation}
Since $e^{- \mathcal D^2 \Delta t/2}e^{\mathcal L \Delta t}=e^{\left(\mathcal L- \mathcal D^2/2\right)  \Delta t}+\mathcal O (\Delta t^2)$, taking the $N\to\infty$ limit gives
\begin{equation}
\overline{\bm \rho(t)}= e^{\left(\mathcal L- \mathcal D^2/2\right)(t-t')}\overline{\bm \rho(t')},
\end{equation}
which implies
\begin{equation}
\frac{d}{dt} \overline{\bm \rho(t)} = \left(\mathcal L-\frac{\mathcal D^2}{2} \right) \overline{\bm \rho(t)}.\label{dephase:lind}
\end{equation}
The noise causes decay of $\overline{\langle\hat\rho_{n_an_bn_a'n_b'}(t)\rangle}$ at a rate $\left[\sqrt{\gamma_a} (n_a-n_a')+\sqrt{\gamma_b} (n_b-n_b')\right]^2/2$.
In the case (\ref{dephase:1}) where the frequency noise on modes $a$ and $b$ are independent, the right-hand side of (\ref{dephase:lind}) is replaced by
$\left(\mathcal L-\mathcal D_a^2/2 -\mathcal D_b^2/{2}\right) \overline{\bm \rho(t)}$ where $\mathcal D_{j,\mu\nu}=\gamma_j\delta_{\mu\nu}(n_j-n_j')$.

The inclusion of dephasing then amounts to adding a decay term on the right hand side of each of the subequations in (\ref{eq:Lindblad-offdiag}) and (\ref{eq:Lindblad-diag})
in Sec.\,\ref{app:spectrum}.
In the case of flux noise (\ref{dephase:2}), the decay rate for $\hubbard_{jj'}$ reads,
\begin{equation}
-\frac{1}{2}\left[\sqrt{\gamma_a}(j\mod 4-j'\mod 4)+\sqrt{\gamma_b}(\lfloor j/4 \rfloor-\lfloor j'/4 \rfloor)\right]^2,
\end{equation}
where the remainder $j\mod4$ and the integer part $\lfloor{j}/4\rfloor$ essentially count the photon numbers $n_a$ and $n_b$, respectively,  in the state $|j\rangle$ defined in Eq.~(\ref{eq:basis})
When the dephasing mechanisms for modes $a$ and $b$ are independent (\ref{dephase:1}), the decay term for $\hubbard_{jj'}$ reads,
\begin{equation}
-\frac{1}{2}\left[\gamma_a(j\mod4-j'\mod4)^2+\gamma_b(\lfloor j/4 \rfloor-\lfloor j'/4 \rfloor)^2\right].
\end{equation} 
The expectation values of the Hubbard operators  are obtained by solving (\ref{eq:Lindblad-diag}) with the added noise terms. In the case of flux noise (\ref{dephase:2}),
the spectral density of
emitted photons is then obtained by adding a diagonal matrix with main diagonal entries
\begin{equation}
\frac{1}{2}\left(4\gamma_a,(\sqrt{\gamma_a}-\sqrt{\gamma_b})^2,\gamma_a,\gamma_a,\gamma_a,(2\sqrt{\gamma_a}-\sqrt{\gamma_b})^2,\gamma_a\right),
\label{dephase:lind1}
\end{equation}
to $\mathcal{M}$ in (\ref{eq:adaggera=}) and solving to leading order in the drive. For independent noise sources (\ref{dephase:1}), the main diagonal reads
\begin{equation}
\frac{1}{2}\left(4\gamma_a,\gamma_a+\gamma_b,\gamma_a,\gamma_a,\gamma_a,4\gamma_a+\gamma_b,\gamma_a\right),
\label{dephase:lind2}
\end{equation} instead.
The resulting equations do not simplify to compact final expressions. The computation can nonetheless be performed efficiently
using symbolic algebra software, such as Mathematica, in order to find an algebraic formula for the purpose of fitting to data.

\begin{figure}[htb]
	\begin{center}
		\includegraphics[width = 1.0\columnwidth]{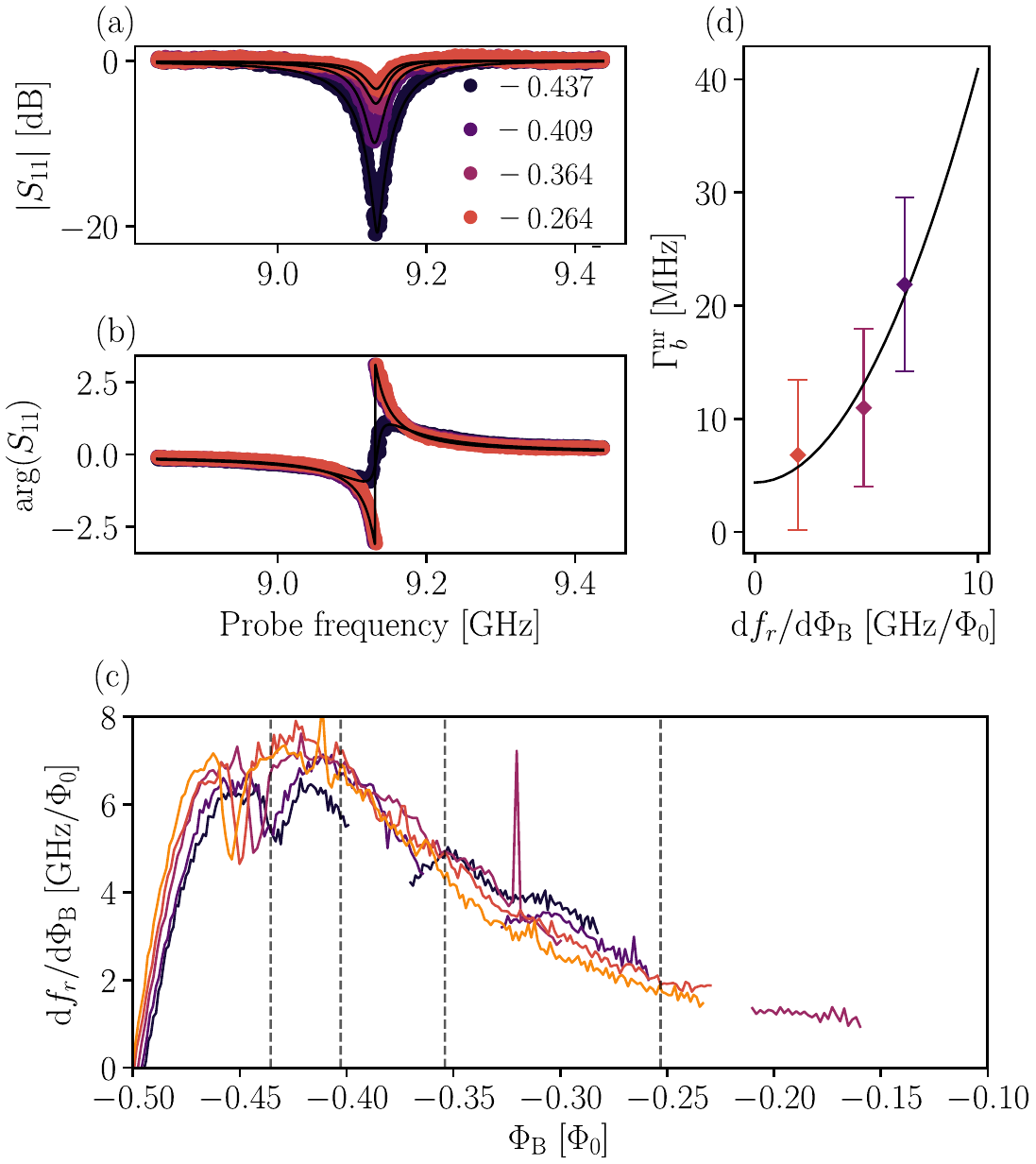}
		\caption{({\bf a} and {\bf b}) Single-tone spectroscopy of mode $b$ for different magnetic flux $\Phi_B$ while keeping the mode resonance frequency constant (thanks to the array
        tunability). The color code is the same as in Figure~\ref{fig:FigAppB1}. The legend provides the value of $\Phi_B/\Phi_0$ (modulo an integer number of flux quanta) at which the data was acquired. Down-conversion is observed at $\Phi_B/\Phi_0=0.437$, but not at the other three values of $\Phi_B/\Phi_0$ for which data is presented. Solid lines are fits to Eq.~(2), used to extract non-radiative losses. ({\bf c}) The slope of frequency as a function of flux (modulo an integer number of flux quanta) on each of the five right-most arcs in Figure~\ref{fig:FigAppB1}. The data in panel ({\bf a}) was obtained at the vertical dashed lines. ({\bf d}) Extracted non-radiative losses versus $df_b/d\Phi_B$ for the three data sets $\Phi_B/\Phi_0=0.264,\,0.364,\,0.409$ where no down-conversion occurs. The solid line is a quadratic fit. The error bars are obtained in the same way as in Figure~\ref{fig:FigAppB2}.}		
        \label{FigAppE2}
	\end{center}
\end{figure}

\subsection{Fitting theory to experiment}
\label{app:fitting}
To compare theoretical predictions for the spectral density of down-converted photons to experimental results, one must know the radiative and non-radiative broadenings 
for modes $a$ and $b$, the transition frequencies $\omega_{10}$, $\omega_{21}$, $\omega_{32}$, and $\omega_b$, and the coupling $g_{1\leftrightarrow 3}$. As mentioned in the main text, we measure down-conversion when $\omega_b=2\pi\times9.13\,\text{GHz}$, and $\omega_{10}=2\pi\times3.16\,\text{GHz}$. We held mode $b$ close to the down-conversion
point, and detuned mode $a$. By then performing single-tone spectroscopy on mode $b$, we determined that it is subject to $\Gamma_b^\text{r}=2\pi\times53\,\text{MHz}$ of radiative broadening and
an additional $22\,\text{MHz}$ of non-radiative broadening. In principle,
this could be due to emission into an unmonitored reservoir (Sec.\,\ref{app:spectrum}), or pure dephasing (Sec.\,\ref{app:dephasing}).

To investigate this further, we measured this non-radiative broadening at various values of flux for which $\omega_b=2\pi\times9.13\,\text{GHz}$, but with mode $a$ detuned, so that broadening is
not dominated by down-conversion. See Figure~\ref{FigAppE2}.
This revealed that the broadening is largest where the slope 
$\partial\omega_b/\partial\Phi_\text{B}$ is largest. We find that this slope dependence
of non-radiative broadening of mode $b$ can be fit by $\Gamma_b^\text{nr,0}+\gamma_b$ with dephasing $\gamma_b$ having a quadratic slope dependence
\begin{equation}  
\gamma_b=A_\Phi \left(\frac{\partial \omega_b}{\partial \Phi_\text{B}}\right)^2,\label{eq:modea}
\end{equation}
with  $A_\Phi=5.8\times 10^{-5} \Phi_0^2/\text{GHz}$, and $\Gamma_b^\text{nr,0}=2\pi\times 4$ MHz due to
sources of broadening other than dephasing by flux noise, such as emission into an unmonitored reservoir. We 
see that at the point where down-conversion is measured, dephasing causes $\sim 18$MHz of broadening, and is thus the dominant source of non-radiative broadening.

Because the fundamental frequency $\omega_{10}$ of mode $a$ lies at the lower edge of the measurement bandwidth, we were unable to perform the same analysis on mode $a$.
From reflection measurements, we could determine $\Gamma_a^\text{r}=2\pi\times84\,\text{MHz}$,
but had to obtain the non-radiative broadening of mode $a$ by an indirect method, that
takes as input, information obtained regarding broadening of mode $b$.
Assuming that non-radiative broadening is dominated by dephasing,
i.e. ignoring a possible small contribution $\Gamma_a^{\text{nr,0}}$, 
we should have for mode $a$
that non-radiative broadening is given by
\begin{equation}  
\gamma_a=A_\Phi \left(\frac{\partial \omega_a}{\partial \Phi_\text{B}}\right)^2,
\end{equation} 
where $A_\Phi$ is the same as in (\ref{eq:modea})
since modes $b$ and $a$ are subjected to the same flux noise. At the point where we measure down-conversion, we have $\partial\omega_a/\partial\Phi_\text{B}=2\pi\times 12.5\,\text{GHz}/\Phi_0$, giving $\gamma_a=2\pi\times57\,\text{MHz}$.

\begin{figure}[H]
\begin{center}
\includegraphics[width = 0.47\textwidth]{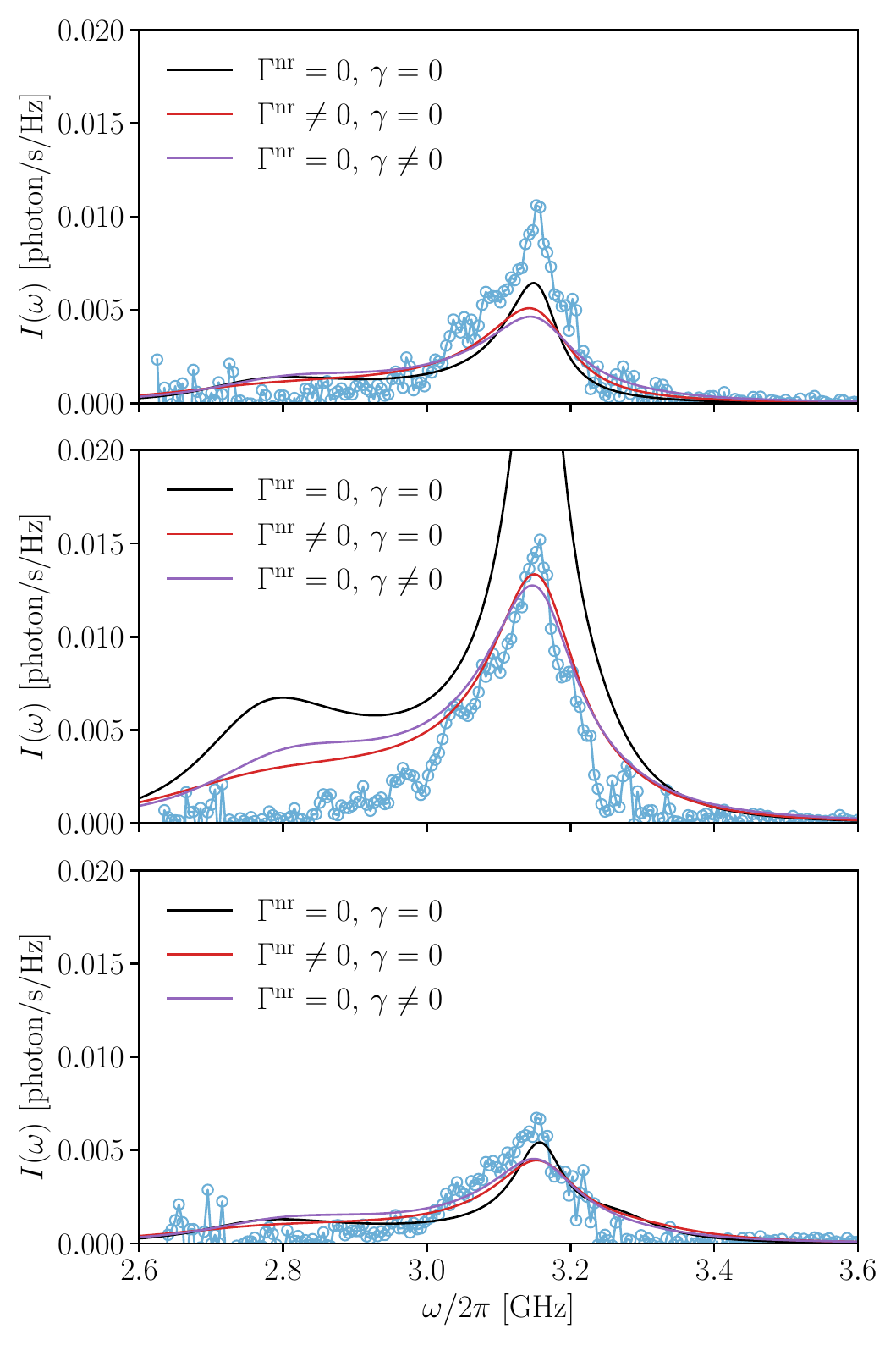}
\caption{Experimental results for the spectral density of down converted photons compared to theory.
The experimental results were all obtained at driving power $-129\:\text{dBm}$. Black curves include only radiative broadening. Red curves include non-radiative broadening $\Gamma_a^\text{nr}=2\pi\times57\,\text{MHz}$ and
$\Gamma_b^\text{nr}=2\pi\times22\,\text{MHz}$ due to emission into an unmonitored reservoir, as in Sec.\,\ref{app:spectrum}. Purple curves on the other hand assume that non-radiative broadening is due entirely to pure dephasing, i.e. purple curves include non-radiative broadening $\gamma_a=2\pi\times57\,\text{MHz}$ and
$\gamma_b=2\pi\times22\,\text{MHz}$. {\bf Top panel:} $\omega_b-\omega_\text{d}=2\pi\times 50\,\text{MHz}$. This data is used to fix the two fitting parameters $\omega_{21}$ and $g_{1\leftrightarrow 3}$.
{\bf Middle panel:} $\omega_b-\omega_\text{d}=0$. The data points and red curve correspond to the $-129\:\text{dBm}$ results of panel Fig.\,\ref{fig3}, panel (c). {\bf Bottom panel:} $\omega_b-\omega_\text{d}=2\pi\times (-50\,\text{MHz})$.}
\label{fig:FigAppI}
\end{center}
\end{figure}

We further tune the flux to obtain the maximum down-conversion signal, which implies that we are on resonance $\omega_{32}+\omega_{21}+\omega_{10}=\omega_b$. This leaves two fitting parameters, $\omega_{21}$ and $g_{1\leftrightarrow 3}$, when we compare the theoretical prediction for the spectral density of down-converted photons to experimental data. We fix these by fitting to data at $\omega_b-\omega_\text{d}=2\pi\times 50\,\text{MHz}$.
To gauge how accurately the theoretical model describes the data, we also present a comparison of theory to experiment at $\omega_b-\omega_\text{d}=0$ and $\omega_b-\omega_\text{d}=-2\pi\times 50\,\text{MHz}$, now without any fitting parameters.

Results are shown in Fig.\,\ref{fig:FigAppI}. We compare three fits. In the first case (black curves) we neglect all non-radiative broadening and pure dephasing, i.e. $\Gamma_a^\text{nr}=\Gamma_b^\text{nr}=\gamma_a=\gamma_b=0$. We see that down-conversion at $\omega_\text{d}=\omega_b$ (middle panel) is predicted rather poorly. In the second case (red curves)
we include $\Gamma_a^\text{nr}=2\pi\times57\,\text{MHz}$  of non-radiative broadening in mode $a$, and
$\Gamma_b^\text{nr}=2\pi\times22\,\text{MHz}$ to mode $b$, which we attribute entirely to emmission in into an unmonitored reservoir, i.e. $\gamma_a=\gamma_b=0$.
In the third case (purple curves)
we ascribe all non-radiative broadening to pure dephasing by flux noise (the assumption 
which we believe most closely corresponds to the actual experimental conditions), i.e. we set
$\gamma_a=2\pi\times57\,\text{MHz}$, 
$\gamma_b^\text{nr}=2\pi\times22\,\text{MHz}$ and $\Gamma_a^\text{nr}=\Gamma_b^\text{nr}=0$.
Comparing theory to experiment at $\omega_b-\omega_\text{d}=0$ and $\omega_b-\omega_\text{d}=-2\pi\times 50\,\text{MHz}$, shows that the inclusion of non-radiative broadening improves the predictive power
of the theory. However, results for broadening by emision into an unmonitored reservoir or by pure
dephasing give similar results. The theory is thus not accurate enough to resolve the distinct non-radiative broadening mechanisms. The fits including non-radiative broadening, that is used in Fig. 3(c) of the main text, yields 
$\omega_{21}\simeq2\pi\times (2.7- 2.8)\,\text{GHz}$ and $g_{1\leftrightarrow 3}\simeq2\pi\times (6-7)\,\text{MHz}$.

\end{document}